\documentclass[api,prb,showpacs,letterpaper,twocolumn,groupedaddress,footinbib,amssymb]{revtex4}

\pdfoutput=1
\usepackage{graphicx}
\usepackage{braket}
\usepackage{color}

\begin{document}

\title{One-dimensional ballistic transport with FLAPW Wannier functions}

\author{Bj\"orn Hardrat$^{1}$}
\email[corresp.\ author: ]{bhardrat@theo-physik.uni-kiel.de}
\author{Nengping Wang$^{2}$}
\author{Frank Freimuth$^{3}$}
\author{Yuriy Mokrousov$^{3}$}
\author{Stefan Heinze$^{1}$}

\affiliation{$^{1}$Institut f\"{u}r Theoretische Physik und
 Astrophysik, Christian-Albrechts-Universit\"{a}t zu Kiel,
 Leibnizstrasse~15, D-24098 Kiel, Germany}
\affiliation{$^{2}$Physics Department, Ningbo University, Fenghua
  Road~818, 315211 Ningbo, People's Republic of China}
\affiliation{$^{3}$Peter Gr\"{u}nberg Institut and Institute for Advanced Simulation,
Forschungszentrum J\"{u}lich and JARA, D-52425 J\"{u}lich, Germany}

\date{\today}
\vspace{1cm}

\begin{abstract}
We present an implementation of the ballistic Landauer-B\"uttiker
transport scheme in one-dimensional systems based on
density functional theory (DFT) calculations within the full-potential linearized
augmented plane-wave (FLAPW) method. In order to calculate the conductance within
the Green's function method we map the electronic structure from the extended states of the
FLAPW calculation to Wannier functions which constitute a minimal localized basis set.
Our approach benefits from the high accuracy of the underlying FLAPW calculations
allowing us to address the complex interplay of structure, magnetism, and spin-orbit coupling
and is ideally suited
to study spin-dependent electronic transport in one-dimensional magnetic nanostructures.
To illustrate our approach
we study ballistic electron transport in non-magnetic
Pt monowires with a single stretched bond including spin-orbit coupling, and
in ferromagnetic Co monowires with different collinear magnetic alignment of the
electrodes with the purpose of analysing the magnetoresistance when going from tunneling 
to the contact regime. We further investigate spin-orbit scattering due to an impurity atom. 
We consider
two configurations: a Co atom in a Pt monowire and vice versa. In both cases, the spin-orbit
induced band mixing leads to a change of the conductance upon switching the magnetization
direction from along the chain axis to perpendicular to it. The main contribution stems
from ballistic spin-scattering for the magnetic Co impurity in the non-magnetic Pt monowire
and for the Pt scatterer in the magnetic Co monowire from the band formed from states with
$d_{xy}$ and $d_{x^2-y^2}$ orbital symmetry. We
quantify this effect by calculating the ballistic anisotropic magnetoresistance which
displays values up to as much as 7\% for ballistic spin-scattering and gigantic values
of around 100\% for the Pt impurity in the Co wire. In addition we show that the presence of
a scatterer can reduce as well as increase the ballistic anisotropic magnetoresistance.
\end{abstract}

\maketitle

\section{Introduction}
\label{sec:Introduction}

With the possibility to perform transport measurements on nano- down to atomic-scale junctions using
mechanically-controllable break-junctions~\cite{KizukaPRB2008} or
scanning tunneling microscopy~\cite{NeelPRL2009,KroegerJPCM2008,TaoPRB2010,ChopraNM2005,CalvoN2009,ZieglerNJP2011}
various fundamental questions on electron transport as well as practical
problems concerning device functionality have arisen.
With shrinking system size the junctions have become considerably smaller than the mean free path of a transmitted electron,
reaching the ballistic transport regime. In this regime various effects such as the geometric arrangement of the atoms,
the chemical composition, the magnetic order, vibrations, correlation effects,
or the magnetic anisotropy can play an important role due to the reduced
coordination number of the participating atoms. In the context of spin-dependent
transport, for example, there is a strong interest in understanding how the spin-valve effect scales to systems
of atomic- or molecular-scale~\cite{ZieglerNJP2011,WulfhekelNN2011}. In nanoscale junctions, new transport effects
can also arise such as the ballistic anisotropic magneto-resistance (BAMR)\cite{VelevPRL2005,SokolovNN2007}.
In order to successfully address such issues a theoretical description
needs to properly take into account the electronic structure of the system which
is typically obtained by first-principles methods based on density functional theory.
The central experimental quantity 
is the measured current versus bias voltage (I-V-curve) or at small bias voltages the conductance.

\begin{figure}
\begin{center}
\centerline{\includegraphics[width=0.40\textwidth,angle=0]{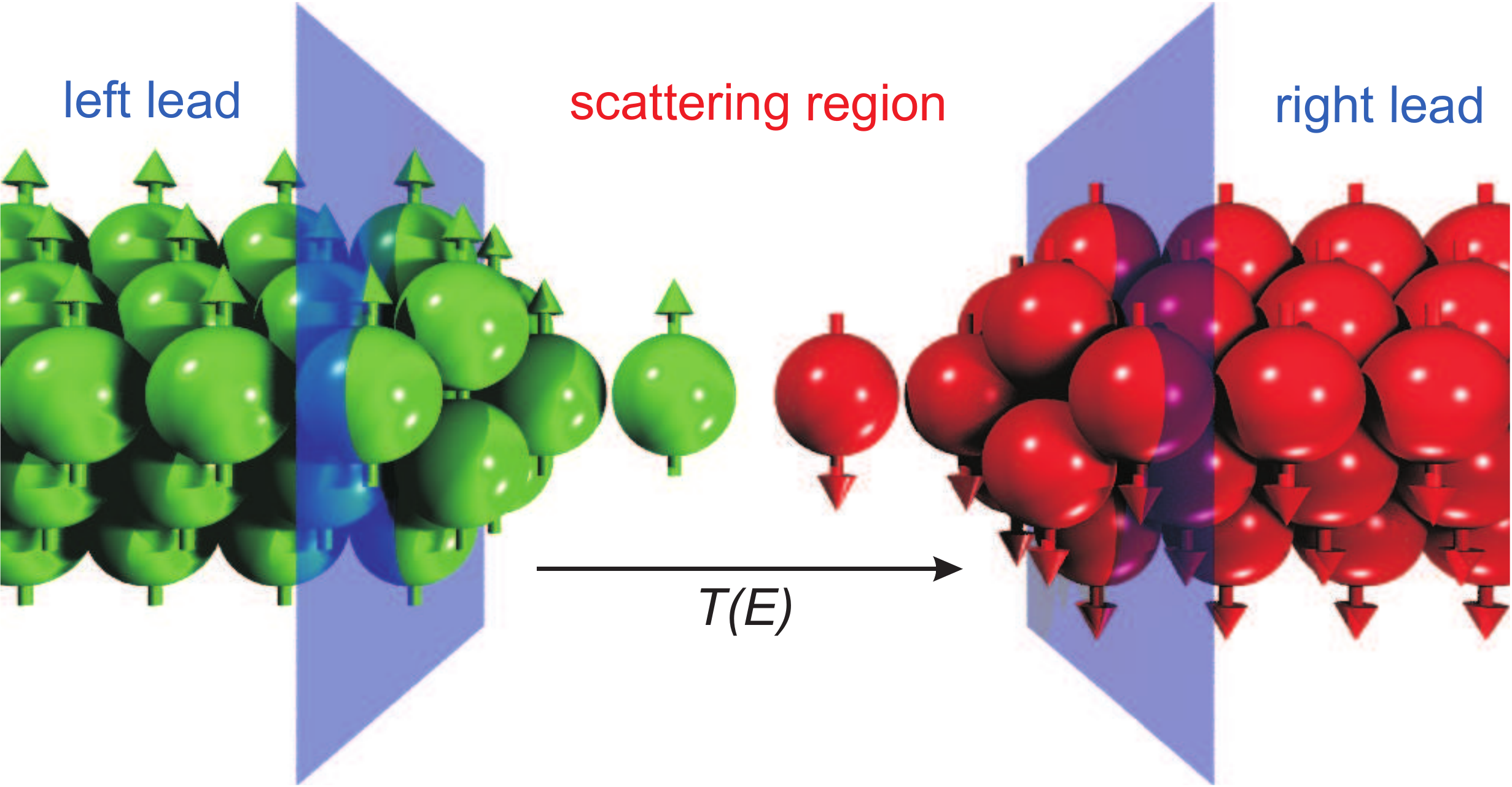}}
\caption{(color online)
Typical geometry of a ballistic transport calculation, consisting of
three different regions (left lead/scattering region/right lead).
Charge carriers with energy $E$ are transmitted through a scattering region with a
transmission probability $T(E)$ from the left lead to the right lead.
The blue planes separate the leads from the scattering region. While
the semi-infinite leads resemble the electronic structure of a periodic
system, the scattering region includes the scatterer as well as the
lead-scatterer contact region.}
\label{fig:TGeo}
\end{center}
\vspace*{-1cm}
\end{figure}

The theoretical method most frequently applied to describe quantum transport in such systems
is the Landauer-B\"uttiker approach in which the junction is divided into a central scattering
region and two leads in thermal equilibrium with contact reservoirs, Fig.~\ref{fig:TGeo}, resulting
in the famous Landauer conductance formula\cite{BuettikerPRB1985}.  Basically two different groups
of techniques have been developed to solve the transport problem: wave-function and Green's function
(GF) based methods which are equivalent in case of non-interacting charge carriers\cite{MeirPRL1992}.
Among the wave function based methods the transmission through such a quasi one-dimensional (1D)
system can be calculated by means of the transfer matrix
method\cite{SautetPRB1988,EmberlyPRB1998,HirosePRL1994,ChoiPRB1999,KobayashiPRB2000},
solving the Lippmann-Schwinger equation\cite{LangPRB1995,DalgleishPRB2005}, or by wave
function matching\cite{KhomyakovPRB2005}.
The GF methods are usually based on Keldysh, Kadanoff and Baym's non-equilibrium Green's
functions (NEGF)\cite{Keldysh,KadanoffBaym}. Beyond the standard non-interacting electron
approach, there has been work incorporating e.g.~inelastic scattering on
vibrations\cite{FrederiksenPRL2004,FrederiksenPRB2007} or treating correlation effects through
self-energies\cite{ThygesenPRB2008}.  An alternative way of calculating quantum transport is
by using the Kubo approach as formulated by Baranger and Stone\cite{BarangerPRB1989}, relating
the current to the dynamical polarization\cite{EversPRB2004,BagretsPRB2006}.

Based on these three general approaches all codes differ in the way the electronic structure
is described. In the first implementations based on density functional theory (DFT), the
electrodes were treated as jellium which were coupled to the scattering
region~\cite{HirosePRL1994,KobayashiPRB2000,LangPRB1995}. Large systems up to devices
can be described using semi-empirical tight-binding methods for the electronic structure
\cite{SautetPRB1988,EmberlyPRB1998,SanvitoPRB1999,DalgleishPRB2005,HaefnerPRB2008} while
approaches using DFT for both the description of the electrodes via self-energies
and the scattering region promise the highest accuracy\cite{NardelliPRB1999,TaylorPRB2001,
NardelliPRB2001,XueCP2002,BrandbygePRB2002,ThygesenPRB2003,CalzolariPRB2004,KePRB2004,
HeurichPRL2002,BagretsPRB2006,DerosaJPCB2001,
PalaciosPRB2001,PalaciosPRB2002,EversPRB2004,ThygesenCP2005,PaulyNJoP2008}.

Among these implementations various DFT methods have been applied. Transport codes based on Green's functions
rely on a localized basis set limiting this approach to basis sets of numerical orbitals
such as Gaussians\cite{XueCP2002,PalaciosPRB2002}, localized
orbitals\cite{TaylorPRB2001,NardelliPRB2001,BrandbygePRB2002,KePRB2004,RochaPRB2006}, or
wavelets\cite{ThygesenPRB2003}. The application of flexible and accurate plane-wave DFT methods
for transport calculations is usually realized in connection
with the scattering approach for the conductance~\cite{SmogunovPRB2004,DalCorsoPRB2006}.
Alternatively, the efficient GF method for the transport calculation can be used
if the extended states in the plane-wave expansion are mapped onto
maximally localized Wannier functions\cite{MarzariPRB1997} (MLWFs). This approach combines
plane-wave calculations with the use of a minimal basis set suitable for quantum transport
calculations\cite{ThygesenPRB2003,CalzolariPRB2004,ThygesenCP2005,Wang2011,autoMLWFsTrans}.
When one is dealing with low-dimensional systems and subtle band structure effects such as spin-orbit coupling
the accuracy of electronic structure description becomes crucial. Therefore, the application of
a highly precise all-electron full-potential linearized augmented plane-wave code is
desirable. To our knowledge no such DFT transport scheme has been reported and only a few codes allow
to incorporate spin-orbit coupling\cite{DalCorsoPRB2006,JacobPRB2008,SmogunovPRB2008,HaefnerPRB2009}.

In this paper we present a method to calculate transport through 1D
nanoscale structures following the Landauer-B\"uttiker approach. The
underlying electronic structure of the studied system is obtained
from DFT within the 1D version\cite{MokrousovPRB2005} of the
full-potential linearized augmented plane-wave method (FLAPW), as
implemented within the {\tt FLEUR} code.\cite{FLEUR} The 1D-FLAPW method
is specifically tailored to treat 1D structures avoiding supercell
calculations: the periodicity is explicitly taken into account only
along the nanostructure's axis ($z$-axis in the following), while the
wavefunctions in the vacuum surrounding the system are forced to obey
an exponential decay in-plane.\cite{MokrousovPRB2005} Since the FLAPW
wavefunctions are intrinsically delocalized in real space we perform a
mapping of the electronic structure of the system onto a set of localized
Wannier functions (WFs), which allows to solve the transport problem in
real space efficiently. The WFs obtained from the FLAPW calculation
(FLAPW WFs)\cite{FreimuthPRB2008,wien2kwannier} provide a minimal localized basis set which describes the
\emph{ab initio} electronic bands within a certain energy window with
high accuracy and allows to efficiently compute the non-equilibrium
Green's function (NEGF) of the system needed to determine its transmission
function $T(E)$. We use and compare two different sets of WFs, namely,
the maximally-localized Wannier functions (MLWFs),\cite{MarzariPRB1997}
which are uniquely defined by fulfilling the condition of maximal
localization in real space, and the so-called first-shot Wannier functions
(FSWFs),\cite{FreimuthPRB2008} being much easier to obtain computationally
and although non-unique, still capable of describing the transport
properties of a system correctly in many cases. A special approximation
we include in our transport scheme is the so-called "locking technique",
which allows to use separately calculated leads and scattering regions and to combine those into one
quantum transport calculation, achieving an
accurate treatment of leads and scattering region at reduced computational
cost.

As a first application we have calculated the electronic structure and the ballistic
transport properties of a non-magnetic Pt monowire with a single stretched
bond in the middle of the chain which acts as a source of scattering. For this rather
simple system we demonstrate the quality of our MLWFs and FSWFs, the locking technique
to obtain the Hamiltonian of the open system, and show the
possibility of decomposing the transmission function in terms of orbital symmetry. We
further investigate the influence of spin-orbit-coupling (SOC) on the transmission of
the Pt wire. We find a substantial change of the conductance of one quantum of conductance
at the Fermi level for a perfect wire due to the strength of SOC in $5d$-transition metals
such as Pt.

In order to include the effect of large spin-polarization we have chosen a ferromagnetic
Co monowire with a single stretched bond, a prototypical magnetic system,
and calculate the magnetoresistance from the conductance in a parallel and antiparallel alignment
of the Co electrodes. We obtain a rapid decrease of the magnetoresistance with the separation
between the two Co monowires which is due to the fast decay of transmission from the highly
spin-polarized localized states of $d_{xz,yz}-$ and $d_{xy,x^2-y^2}-$symmetry.

Finally, we have studied scattering from a single impurity atom in a monowire
due to SOC. We have chosen two configurations: (i) a non-magnetic Pt atom in a
ferromagnetic Co wire and (ii) a magnetic Co atom in a non-magnetic Pt wire. In
both cases, we have compared the conductance obtained in the scalar-relativistic
approximation and upon including SOC. We find a strong influence of SOC on the
transmission due to the induced splitting of bands. In addition, the conductance
depends sensitively on the magnetization direction in the system being either
along the wire axis or perpendicular to it. While in case (ii) the resulting ballistic anisotropic
magnetoresistance displays values of 7 \% due to spin-orbit interaction mediated
scattering into both spin-channels for the symmetry breaking out-of-chain quantization axis,
in case (i) the values of BAMR reach as much as 100\%, reflecting the giant value
of the ballistic anisotropic magnetoresistance of the pure Co chain.

The paper is organized as follows. In Sec.~\ref{sec:Method} we describe the
theoretical basis of our approach to calculate the conductance and introduce
the key quantities. In particular, the Green's function method is applied to
obtain the transmission function and the conductance. The mapping of the
electronic structure from the FLAPW method to a localized basis set is
accomplished via Wannier functions. The construction of the Hamiltonian for
the open quantum system is described. In Sec. \ref{sec:Results} we present the
first applications of our new transport code to several typical systems of
interest. We begin with the conductance for a non-magnetic Pt wire
with a single broken bond and study the transmission as a function of
bond length and upon including spin-orbit coupling. Then
the magnetoresistance of Co monowires with a single elongated bond
is discussed. Finally, the effect of spin-orbit scattering is illustrated
by two examples: a Pt monowire with
a single magnetic Co atom and a Co monowire with a single Pt atom.
A summary and conclusions are given in Sec.~\ref{sec:Summary}.


\section{Method}
\label{sec:Method}
\subsection{General transport problem}

We describe the transport properties of the system within the
Landauer-B\"uttiker approach, dividing it into three different
regions: two semi-infinite leads (left, $L$, and right, $R$)
and the scattering region ($S$), which includes the actual scatterer
as well as the lead-scatterer contact region, in which the effect
of the scatterer on the properties of the leads ideally decays
such that their electronic structure can be considered perfect
and unperturbed inside the $L$ and $R$ regions. Assuming that
the interaction between the left and right leads can be neglected
the tight-binding Hamiltonian of our system corresponding to such a
structural division has the following form:
\begin{equation}\label{index1}
\label{OpenSys}
\mathbf H=\left(
\begin{array}{lll}
\mathbf H_{L}&\mathbf H_{LS}^\dagger&0\\
\mathbf H_{LS}&\mathbf H_{S}&\mathbf H_{SR}\\
0&\mathbf H_{SR}^\dagger&\mathbf H_{R}
\end{array}\right),
\end{equation}
where $\mathbf H_{L/R}$ is the semi-infinite Hamiltonian of the left/right
lead, while $\mathbf H_{LS/SR}$ describes the coupling of the scattering
region to the leads and $\mathbf H_{S}$ is the Hamiltonian of the scattering
region. Due to the semi-infiniteness of the leads the dimension
of the Hamiltonian, Eq.~(\ref{index1}), is infinite, which presents
a conceptual computational problem. An efficient method to deal
with that, applicable to any system of the type depicted in Fig.~1,
which can be described with a real-space tight-binding Hamiltonian
of the type of Eq.~(1), has been developed.\cite{DattaBook} This
method is based on the non-equilibrium Green's function (NEGF)
formalism, 
which treats the scattering region and the semi-infinite
leads on equal footing, describes extractions, re-injections and
excitations of electrons in the system and solves the problem of
the semi-infinite leads by introducing finite-dimensional self-energies
$\mathbf \Sigma_{L/R}$ which include the true lead's effect on the scattering
process. Within the NEGF formalism the system is described by means
of the retarded Green's function:
\begin{equation}
\label{totalGF}
\mathbf G(E)=[(E+i\epsilon)\mathbf I -\mathbf H]^{-1},
\end{equation}
where $\mathbf I$ denotes the unity matrix of the dimension of
$\mathbf H$. By neglecting at first the coupling of the leads
to the scattering region and regarding just the first few layers
of the leads which are actually interacting with the scattering region,
it is possible to replace the leads Green's function by their surface Green's
functions $\mathbf{g}_{L/R}(E)$.\cite{DattaETiMSBook}
This can be derived by rewriting the lead's Hamiltonian in a block
diagonal form using square matrices $\mathbf{h}_{L/R}$ and $\mathbf{h}_{LL/RR}$
of the same dimension as the surface Green's function:
\begin{equation}
\label{leadH}
\mathbf H_L=\left(\begin{array}{cccc}
\ddots&&&\mathbf{0}\\&\mathbf{h}_L&\mathbf{h}_{LL}^\dagger&\\&\mathbf{h}_{LL}&
\mathbf{h}_L&\mathbf{h}_{LL}^\dagger\\\mathbf{0}&&\mathbf{h}_{LL}&\mathbf{h}_L
\end{array}\right).
\end{equation}

Based on this description of the leads, the surface Green's
function $\mathbf{g}_{L/R}(E)$ can be determined iteratively starting from
\begin{equation}
\label{surfaceGF}
\mathbf{g}^{[0]}_{L/R}(E)=[(E+i\epsilon)\mathbf I_{L/R}-\mathbf{h}_{L/R}]^{-1},
\end{equation}

with $\mathbf{g}^{[0]}_{L/R}(E)$ being a square matrix with the dimension of
interacting orbitals at the leads' surfaces. The expression~(\ref{surfaceGF})
can be converged to the surface Green's function by recursively incorporating the
inter-layer interaction sub-matrices $\mathbf{h}_{LL/RR}$
with an efficient recursive scheme.\cite{GuineaPRB1983}

By reintroducing the coupling of the scattering region to the
leads as a perturbation to the system, the Green's function $\mathbf G_S(E)$
of the scattering region can be obtained from the unperturbed Green's function
of the scattering region by the Dyson equation
\begin{equation}
\label{scatteringGF}
\mathbf G_{S}(E)=[E\mathbf I_S-\mathbf H_{S}-\mathbf H_{LS}^\dagger 
\mathbf{g}_{L}\mathbf H_{LS}-\mathbf H_{SR}^\dagger \mathbf{g}_{R}\mathbf H_{SR}]^{-1}.
\end{equation}
The whole effect of the semi-infinite leads on the conductor
can be then expressed by the leads' self-energies $\mathbf \Sigma_{L/R}(E)$,
which incorporate the surface Green's function $\mathbf{g}_{L/R}(E)$ and
the now finite-sized coupling matrices
$\mathbf H_{LS/SR}$, adapted to the size of the surface Green's functions:
\begin{equation}
\mathbf \Sigma_{L/R}(E)=\mathbf H_{LS/SR}^\dagger \mathbf{g}_{L/R}(E)\mathbf H_{LS/SR}.
\end{equation}
The self-energies are obviously finite-sized matrices of the
dimension of $\mathbf H_S$. The self-energies are related to the
broadening matrices $\Gamma$:
\begin{equation}
\mathbf \Gamma_{L/R}(E)=i[\mathbf \Sigma_{L/R}(E)-\mathbf \Sigma_{L/R}^\dagger(E)],
\end{equation}
which describe the effect of broadening of the states in
the scattering region caused by the presence of the leads
as well as the transfer rates of charge carriers from the
leads into the scattering region.
The incorporation of the non-Hermitian self-energies changes the nature
of the description from the static steady-state picture of the
open system to a dynamic transport scheme, responding to an incoming
charge carrier with the energy $E$.
Based on these quantities
the transmission function $T(E)$, describing the probability
of charge carriers originating from one lead to be transmitted
to the other lead, can be expressed in the following way:
\begin{equation}
\label{Transmission}
{T(E)=\rm Tr}[\mathbf G_S(E)\mathbf \Gamma_L(E)\mathbf G_S^\dagger(E)\mathbf \Gamma_R(E)].
\end{equation}
The current, being a natural observable in a quantum transport
measurement, can then be calculated from the Landauer formula:
\begin{equation}
I=\frac{e}{h}\int dE\:T(E)[f_L(E)-f_ R(E)],
\end{equation}
where $f_{L/R}$ are the occupation functions of the leads. The
expression for the conductance then reads:
\begin{equation}
G(E)=\frac{e^2}{h}T(E).
\end{equation}
In case of perfect transmission, $T(E)=1$, this results in
the well-known conductance quantum
\begin{equation}
G_0=\frac{2e^2}{h}.
\end{equation}
for a single, spin-degenerate band.

\subsection{From FLAPW states to localized Wannier functions}
\label{FLAPWtoWFs}

The aim of the approach introduced here is to combine the accuracy and
speed of state-of-the-art DFT electronic
structure calculations based on the one-dimensional version of the FLAPW
method as implemented in
the \texttt{FLEUR} code,\cite{MokrousovPRB2005} and the capability of
the NEGF-formalism described above to treat the whole transport
problem in an efficient way. Especially for transport phenomena driven
by magnetism or spin-orbit coupling (SOC) a precise description of the electronic
structure is necessary.
Typical systems currently under
scrutiny in experiment include geometries with a low coordination
number which favors magnetism and gives rise to strong SOC due to
unquenching of the orbital moment.\cite{NeelPRL2009}

The major problem in combining an LAPW or a plane-wave based electronic
structure method with the real-space transport schemes lies in the
fact that normally several hundreds of delocalized basis functions per
atom are used in such codes in order to achieve the required accuracy.
In our implementation we use the machinery of Wannier functions (WFs),
constructed out of FLAPW wavefunctions,\cite{FreimuthPRB2008} which
proved to be an efficient connection between the two, conceptually
independent, computational methods. The main advantage in such a
"link" can be attributed to two factors: (i) using the gauge freedom of
Wannier functions they can be enforced to be rather localized in real
space, and (ii) an "exact" mapping of the {\it ab initio} Hamiltonian
onto a tight-binding representation with WFs as a localized
orthonormal basis set can be achieved.\cite{SouzaPRB2001}

Having at hand the converged Bloch wavefunctions
$\psi_{m\mathbf{k}}$ for a set of bands $m\le M$ calculated
on a uniform mesh of $\mathcal{N}$ $k$-points, the orthonormal set of Wannier
functions can be obtained via the following transformation\cite{WannierPR1937}:
\begin{equation}
\label{Umn}
\Ket{W_{\mathbf Rn}}=\frac{1}{\mathcal{N}}\sum_{\mathbf k}e^{-i\mathbf
  k\cdot\mathbf R}\sum_{m=1}^MU_{mn}^{\mathbf k}\Ket{\psi_{\mathbf km}},
\end{equation}
where the number of WFs $N$ should be smaller than or equal to $M$.
The gauge freedom
of WFs manifests itself in that the matrices $U_{mn}^{\mathbf k}$ (in the
following, $\mathbf U$-matrices) can in
principle be arbitrary. In the case when $N=M$ and the group of bands from which we
are extracting the WFs from is isolated from other bands, the $\mathbf U$-matrices
are unitary at each $k$-point. Imposing the constraint of maximal
localization of WFs in real space determines the set of $\mathbf U$-matrices
up to a common global phase, and the corresponding set of WFs is
called the maximally-localized Wannier functions (MLWFs).\cite{MarzariPRB1997}
For the whole procedure of maximal localization we use the \texttt{Wannier90} code.\cite{wannier90}

The criterion for the localization of WFs is the smallness of their
spread.\cite{MarzariPRB1997} The process of the spread minimization
constitutes an iterative process at the end of which the $\mathbf U$-matrices
corresponding to the MLWFs are obtained. This minimization procedure
requires as a starting point a certain initial guess for the set of the MLWFs.
In order to construct this set, one chooses certain localized orbitals
$\Ket{g_n}$, which are projected onto the subspace of wavefunctions $\Ket{\psi_{\mathbf km}}$:
\begin{equation}
\Ket{\phi_{\mathbf kn}}=\sum_m\Ket{\psi_{\mathbf km}}\Braket{\psi_{\mathbf km}| g_n},
\end{equation}
and then orthonormalized:
\begin{equation}
\Ket{\tilde{\psi}_{\mathbf kn}}=\sum_m\left((\mathbf S^{(\mathbf k)})\right)
^{-\frac{1}{2}}\Ket{\phi_{\mathbf km}},
\end{equation}
with the overlap matrix $S^{(\mathbf k)}_{mn}=\Braket{\phi_{\mathbf km}|\phi_{\mathbf kn}}$,
after which the starting WFs can be generated:
\begin{equation}
\Ket{W_{\mathbf Rn}}=\frac{1}{\mathcal{N}}\sum_{\mathbf k}e^{-i\mathbf
  k\cdot\mathbf R}\Ket{\tilde{\psi}_{\mathbf kn}}.
\end{equation}
This orthonormal set of Wannier orbitals we will call in the following
the first-shot WFs (FSWFs).

The FSWFs are not unique in the sense that they strongly depend on the choice
of the localized orbitals $g_n$. In many cases, however, especially when MLWFs
are well-localized around atoms as in the case of certain $d$-orbitals in
most of transition-metals and transition-metal oxides,\cite{anisimov} the difference between
the FSWFs, originated from the localized $d$-orbitals, and the corresponding MLWFs
is rather small. This allows to spare the computational
time needed for the minimization of the spread, and immediately construct,~e.g.,
the needed effective Hamiltonians in terms of FSWFs. Examples, when there is a
substantial difference between the FSWFs and MLWFs, include
orbitals for which the centers of the WFs do not coincide with the centers of atoms.
In the following we will analyze in detail the difference in transport properties
calculated with MLWFs and FSWFs, both in the case when there is little difference
between the two sets of WFs and when the difference between them is significant.

\subsection{Construction of the Hamiltonian in real space}
\label{subsection:construction}

In terms of the FLAPW basis functions the Hamiltonian can be written as
\begin{equation}
\label{Ham-BF}
\mathbf{H}_{\mathrm{FLAPW}}=\frac{1}{\mathcal{N}}\sum_{m\mathbf{k}}\varepsilon_{m}(\mathbf{k})
\Ket{\psi_{m\mathbf{k}}}\Bra{\psi_{m\mathbf{k}}},
\end{equation}
while in terms of WFs the equivalent expression is
\label{Hamiltonian}
\begin{equation}
\mathbf{H}_{\mathrm{WFs}}=\sum_{n\mathbf{R}_1}\sum_{n'\mathbf{R}_2}
H_{n,n'}(\mathbf{R}_1-\mathbf{R}_2)
\Ket{W_{n\mathbf{R}_1}}\Bra{W_{n'\mathbf{R}_2}},
\end{equation}
where
\begin{equation}
\label{Ham-hopp}
H_{n,n'}(\mathbf{R}_1-\mathbf{R}_2)=\Braket{W_{n\mathbf{R}_1}|\mathbf{H}_{\mathrm{FLAPW}}|
W_{n'\mathbf{R}_2}}
\end{equation}
are the hopping integrals between the $n$ and $n'$ Wannier orbitals at
sites $\mathbf{R}_1$ and $\mathbf{R}_2$. By substituting Eq.~(\ref{Ham-BF})
into Eq.~(\ref{Ham-hopp}) we find:
\begin{equation}
\label{HoppingElements}
H_{n,n^\prime}(\mathbf{R}_1-\mathbf{R}_2)=\frac{1}{\mathcal{N}}\sum_{m,\mathbf k}
\varepsilon_{m}(\mathbf{k})\Braket{W_{n\mathbf{R}_1}|\psi_{m\mathbf{k}}}
\Braket{\psi_{m\mathbf{k}}|W_{n'\mathbf{R}_2}}
\end{equation}
Thus the real-space representation of the Hamiltonian in terms of localized
Wannier functions can be derived from the knowledge of the eigenvalues and
wavefunctions of the system. In respect to WFs, for efficient evaluation
of Eqn.~(\ref{HoppingElements}) only knowledge of the $\mathbf U$-matrices is required.
\cite{FreimuthPRB2008} The correspondence in the eigenspectrum between the
constructed Hamiltonian in terms of WFs and the Hamiltonian in terms of
eigenfunctions is {\it exact} on the grid of $k$-points used for the WFs
construction, and for this reason the set of WFs is sometimes referred to as
{\it exact} basis set, or, the tight-binding basis set of {\it ab initio}
accuracy.
(This is only valid within the frozen inner window of disentangled systems \cite{disentanglement}).

\begin{figure}
\begin{center}
\centerline{\includegraphics[width=0.40\textwidth]{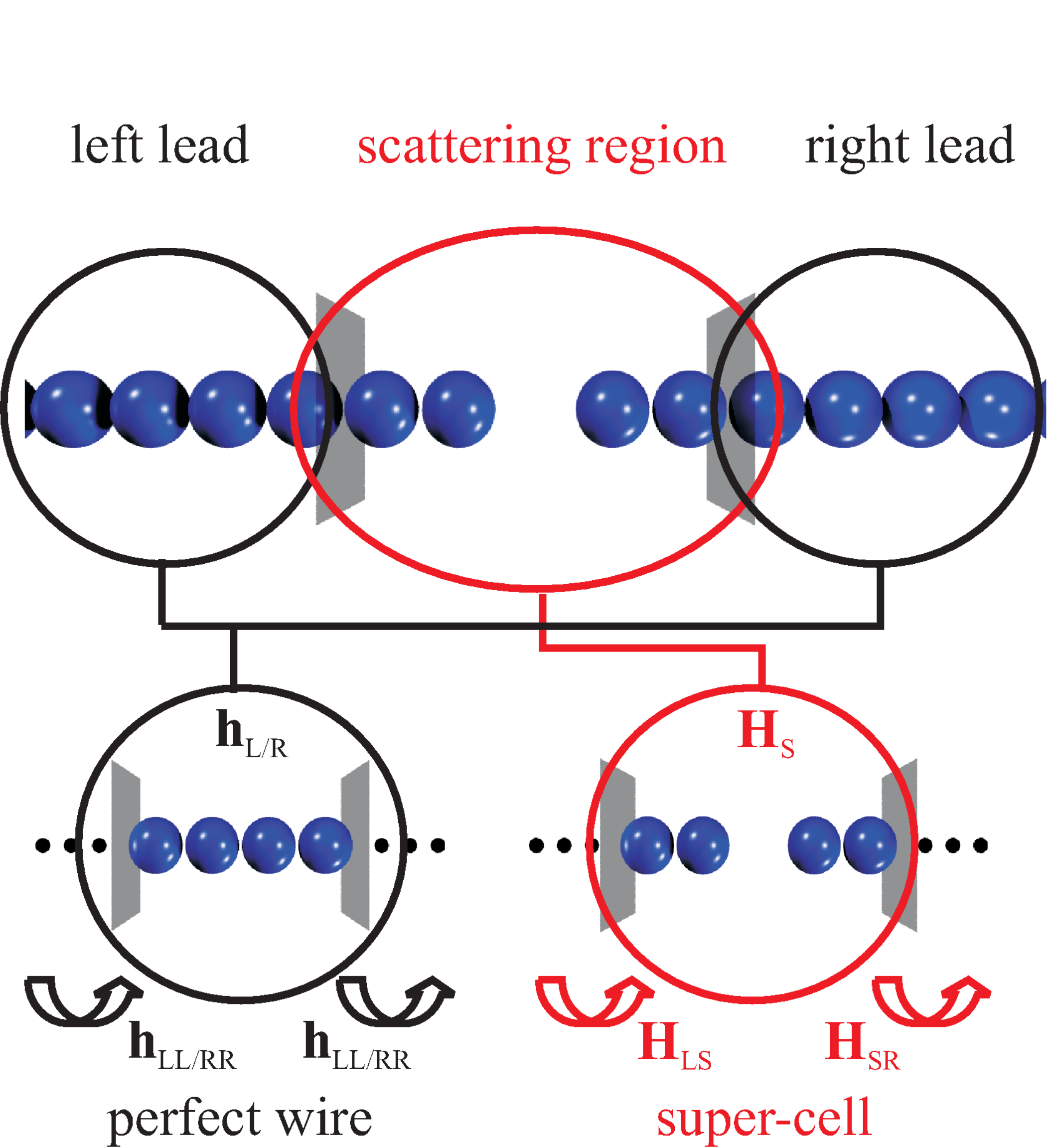}}
\caption{(color online)
Schematic sketch of a ballistic transport calculation based on a WFs
tight-binding Hamiltonian. The leads are described by perfect wires to exclude
spurious deviations from their exact electronic structure. Their semi-infinite
structure is constructed from the Hamiltonians of principal layers $\mathbf{h}_{L/R}$ and the
interaction matrices $\mathbf{h}_{LL/RR}$ between two principal layers.
The scattering region is described by the Hamiltonian $\mathbf H_S$ and coupled to the leads
by the interaction matrices $\mathbf H_{LR/SR}$,
extracted from a supercell calculation. The supercell has to be large enough to reproduce
the lead-scatterer contact with desired accuracy, usually larger than sketched here.}
\label{fig:locking}
\end{center}
\vspace*{-1cm}
\end{figure}

According to Eqns.~(\ref{surfaceGF}) to (\ref{Transmission}), in order to calculate
the transmission function only the $\mathbf{h}_{L/R}$, $\mathbf{h}_{LL/RR}$, $\mathbf H_{LS/SR}$ and $\mathbf H_S$ parts
of the Hamiltonian are needed. Given a {\tt FLEUR} electronic structure calculation it
is necessary to construct these parts of the Hamiltonian from the resulting WFs hopping
elements (Eqn.~\ref{HoppingElements}). We focus on the correct treatment of the
scattering region (see Fig.~(\ref{fig:locking})).

After determination of the atoms belonging to the scattering region, it
is possible to write down the preliminary result for $\mathbf H_S$, based on Eqn.~(\ref{HoppingElements}),
\begin{equation}
\label{H_S}
\mathbf H_S=\sum_{i,n}\sum_{j,m}
H_{n,m}(\mathbf{R}_i-\mathbf{R}_j)\Ket{W_{n\mathbf{R}_i}}\Bra{W_{m\mathbf{R}_j}}
\end{equation}
where $i$ and $j$ determine the atom and $n$ and $m$ the inherent WFs.

Due to the real-space decay of the WFs the corresponding hopping
matrix elements $H_{n,n^\prime}(\mathbf{R}_i-\mathbf{R}_j)$ also decay as the
distance in real space between the Wannier functions $|\mathbf{R}_i-\mathbf{R}_j|$
is increasing.
For an efficient use of the real-space WFs Hamiltonian within the transport scheme
described above it is necessary to keep its matrix elements only up to a certain
number of nearest neighbors (n.n.), setting the rest of the elements to zero.
As a result of this procedure the Hamiltonian matrix becomes sparse, which allows
for a computationally inexpensive treatment. For a given number of n.n.,
the {\it quality} of the sparse Hamiltonian depends on the degree of localization 
of the WFs. Here, by quality of the Hamiltonian we mean the correspondence between
its eigenvalue spectrum to that obtained from {\it ab initio}, or, in the sense of
transport, how well-converged the transmission function $T(E)$ is with respect to
the number of n.n. In this respect, in the following we compare and analyze the
results obtained with MLWFs and FSWFs, which display different localization properties.

One way to deal with the exponential decay in Eqn.~(\ref{H_S}) would be to manually
eliminate all matrix elements beyond a certain n.n.. We propose here a 
flexible scheme, minimizing this effort by dividing the scattering region into principal layers
$\mathbf{h}_l$, $l=1,\ldots,s$ and interaction matrices $\mathbf{h}_{l,l+1}$ between neighboring layers:
\begin{equation}
\label{Matrix_H_S}
\mathbf H_S=\left(\begin{array}{cccc}
\mathbf{h}_{1}&\mathbf{h}_{12}^\dagger&&\mathbf{0}\\\mathbf{h}_{12}&\ddots&\ddots&\\&\ddots&\mathbf{h}_{s-1}&\mathbf{h}_{(s-1)s}^\dagger\\
\mathbf{0}&&\mathbf{h}_{(s-1)s}&\mathbf{h}_{s}
\end{array}\right)
\end{equation}
The sub-matrices are set up as Eqn.~(\ref{H_S}). For the on-site
matrices, $\mathbf{h}_l$, the indices $i$ and $j$ are restricted to atoms from the given layer $l$. For the
interaction matrices $\mathbf{h}_{l,l+1}$ the index $i$ is restricted to atoms from layer $l$ and the index
$j$ to atoms of the neigboring layer $l+1$..

While still capable of describing the system in terms of Eqn.~(\ref{H_S}) (with $s=1$),
the principal layers can optimally contain the number of atoms effectively interacting,
reducing the number of neglected hoppings. Typically these principal layers are
chosen to contain the same number of atoms as the principal layers of the leads
resulting in the same aproximation in terms of n.n. for both regions
and thereby avoids inconsistencies in the transort calculations.
Furthermore this scheme allows
possible future extensions, such as e.g. a combination of separately calculated scatterers into
one scattering region. 

Knowing the Hamiltonian $\mathbf H_S$ of the scattering region, it is necessary to determine the
coupling of the scattering region to the leads. Since the Hamiltonians of both leads and the
scattering region are partitioned into principal layers (see Eqn.~(\ref{leadH}) and Eqn.~(\ref{Matrix_H_S})),
we only need to find the hopping elements between the adjacent layers. Interactions between non-neighboring layers are
neglected by construction. The non-zero elements of $\mathbf H_{LS}$ can now be extracted
from the supercell calculation (see Fig.~\ref{fig:locking}) as
\begin{equation}
\label{H_LS}
\mathbf H_{LS}=\sum_{i,n}\sum_{j,m}
H_{n,m}(\mathbf{R}_{i}-\mathbf{R}_{j})\Ket{W_{n\mathbf{R}_i}}\Bra{W_{m\mathbf{R}_j}},
\end{equation}
where the index $i$ runs over the atoms of the principal layer $1$ of $\mathbf H_S$ and the index $j$ runs over
the principal layer of the left lead.
$\mathbf H_{SR}$ can be constructed analogously.
To prevent a
significant systematical error, it is necessary to make the original supercell large enough to
screen an unphysical inter-unit cell interaction.

Finally, only the Hamiltonians for the leads are missing. Ideally, the calculated unit cell
should be large enough in order to reproduce the properties of the bulk material far away 
from the scatterer and 
thus, the lead Hamiltonian can be extracted directly from the supercell calculations in
a straight-forward manner. Owing to the significant computational burden, it is, however, 
hardly feasible to apply this approach to large and complex systems while keeping at the same time 
the accuracy necessary to capture the main energy scales of the phenomena studied. The technique we use to
overcome this problem, particularly prominent for the FLAPW method with its complicated
basis set, is discussed in the following section.

Up to now no comments have been made concerning the way magnetic systems and the effect
of SOC are treated.
For magnetic systems the majority and minority spin channels can be regarded separately,
resulting in two independent calculations of the transmission function for spin-up and
spin-down channels. In the presence of SOC the whole methodology holds considering that
both spin channels have to be treated together, thus resulting in twice 
the number of WFs used simultaneously to solve the transport problem.

\subsection{Locking technique}
\label{subsection:locking}

The accurate treatment of the leads within the approach described above constitutes a
considerable challenge. Taken from a self-consistent supercell electronic structure
calculation as they are, the sub-matrices $\mathbf{h}_{L/R}$ and $\mathbf{h}_{LL/RR}$ will contain deviations
from "ideal"-lead matrix elements in a large vicinity of the scattering region. While
some of these deviations are definitely physical in their origin due to a large
decay length of 1D charge perturbations caused by the scatterer, the rest of them will
be a spurious artefact of the supercell approach owing to the fact that the leads
as calculated are not intrinsically semi-infinite. This presents a considerable problem
in particular when the leads have to be described with Hamiltonians beyond the $1^{\rm{st}}$
n.n. In this case to describe the semi-infinite leads precisely one would have to go to huge
supercells so that the $A$ atoms in the supercell describing the lead would be exactly
identical, with $A$ being the number of atoms in one principal layer (see Eqn.~(\ref{leadH})).
We found that condition impossible to achieve for non-trivial systems.
Another approach of constructing a lead beyond $1^{\rm{st}}$ n.n. artificially from the outmost
atoms of the scattering region by periodically expanding it is flawed, too, due to the
unknown unperturbed hopping matrix elements beyond $1^{\rm{st}}$ n.n..
This is a serious problem, since the lead has to be described as precisely as possible to
prevent a huge systematic error.

The basic idea to work around this problem is as simple as effective, namely matching
the supercell hopping matrix elements to those of the true leads. Within this so-called "locking"
technique, the leads are replaced by the perfect wires, providing correct
self-energies and Fermi levels of the true infinite periodic system, while the supercell
size is chosen large enough to describe the lead-scatterer interface region sufficiently
well, see Fig.~\ref{fig:locking}. In our transport approach this means that different
parts of the Hamiltonian (Eqs.~(\ref{H_S})~and~(\ref{H_LS})) are extracted from
two different DFT calculations~\cite{LeePRL2005,RochaPRL2008,ShellyEPL2011}: 
the $\mathbf H_S$ and $\mathbf H_{LS/SR}$ coupling matrices are taken
from the supercell calculation describing the scattering region, while the $\mathbf{h}_{L/R}$
and $\mathbf{h}_{LL/RR}$ sub-matrices (needed in Eqn.~(\ref{leadH})) are taken from the calculation
for the perfect leads. $\mathbf{h}_{L/R}$ and $\mathbf{h}_{LL/RR}$ can be determined similar to the principal
layers $\mathbf{h}_l$ and $\mathbf{h}_{l,l+1}$ of $\mathbf H_S$ (Eqns.~(\ref{H_S})~and~(\ref{Matrix_H_S}))
with the principal layer $l$ and the neighboring identical layer $l+1$. To achieve
matching Fermi levels for lead and supercell calculations,
it is additionally necessary to align the diagonal elements
of the matrices $\mathbf{h}_{L/R}$ (Eqn.~\ref{leadH}) and $\mathbf H_S$ (Eqn.~\ref{Matrix_H_S}).

\section{$\rm{Pt}$ monowires}
\label{sec:Results}

In the following sections we present a few instructive applications which illustrate the quality
and possibilities of our FLAPW-WF based approach to obtain the conductance in one-dimensional
magnetic systems within the Landauer coherent transport method.  In this section, we focus on
Pt monowires which possess a single stretched bond that acts as a scattering potential for electrons.
Starting from the construction of the WFs and the tight-binding like Hamiltonian, we discuss
the transmission function and its decomposition in eigenchannels. Our results further demonstrate
the applicability of the locking technique described above. Finally, we include spin-orbit
coupling in our calculations and show that the obtained transmission compares well
with that calculated based on the scattering approach in combination with a
pseudopotential method for the electronic structure\cite{DalCorsoPRB2006}.

In order to calculate the conductance within the approach described in the
previous sections we need to perform two separate DFT calculations and
subsequent Wannierizations for every system: (i) a calculation for the
semi-infinite electrode and (ii) a supercell calculation which includes
the scattering center. From the latter, we determine the hopping matrix
elements for the coupling to the leads. For the monowires considered in
the following the Hamiltonian of the semi-infinite electrode can be
obtained from a calculation with one atom in the unit cell. For the
scattering region, we have used supercells of different size as
described in the computational details section in the appendix.

\subsection{Bandstructure and hoppings}
\label{subsec:FSWFs_versus_MLWFs}

\begin{figure}
\begin{center}
\centerline{\includegraphics[width=0.5\textwidth,angle=270]{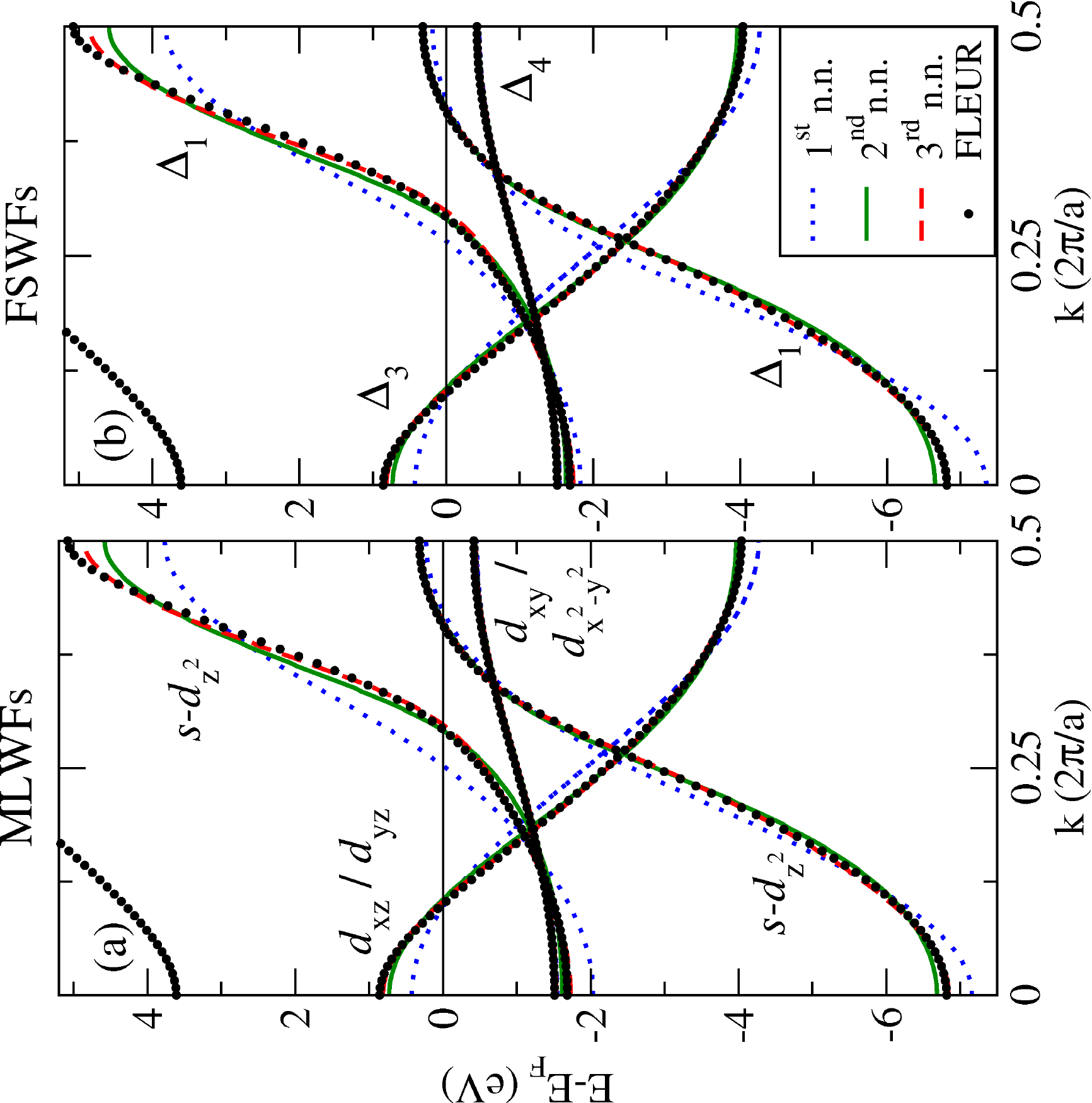}}
\caption{(color online)
Comparison of the Pt monowire bandstructure in SR approximation calculated within DFT
(big black dots) and obtained from the WF Hamiltonian based on (a) MLWFs and (b) FSWFs
considering a limited number of nearest neighbors. In (a) the orbital character of the
states is given and in (b) the bands are denoted according
to their symmetry with respect to the chain geometry.}
\label{fig:MLWFs_vs_FSWFs_BS}
\end{center}
\vspace*{-1cm}
\end{figure}

Before proceeding into the discussion of the transmission, it is insightful to
examine the localization properties
of typical MLWFs and FSWFs which we use for our transport calculations. While the unique
MLWFs are rather well localized in real space, this is not necessarily
the case for the FSWFs, which strongly depend on the choice of the initial
orbitals. If the trial orbitals do not differ very much from the final result of
the localization procedure the difference in spread between the MLWFs and the FSWFs can be small.

For transition-metal monowires this is the case for the localized $d$-orbitals of $\Delta_3$
symmetry ($d_{xz}$ and $d_{yz}$) and of $\Delta_4$ symmetry ($d_{xy}$ and $d_{x^2-y^2}$).
Taking an infinite periodic Pt monoatomic chain with an interatomic spacing of 4.48 bohr
as an example we calculate the spreads of the $\Delta_3$ and $\Delta_4$ MLWFs
to be 3.70 bohr$^2$ and 2.22 bohr$^2$, respectively. The calculated spreads of
the FSWFs, constructed with solutions of the radial equation for the
actual potential obtained from the first-principles calculation~\cite{FreimuthPRB2008},
are indistinguishable from the former.

The situation is completely different, however, for the FSWFs constructed from the
$s$- and $d_{z^2}$-like trial orbitals. In this case the difference
in spread between the resulting FSWFs and the $\Delta_1$-like MLWFs is remarkable.
While values of 2.89 and 6.20 bohr$^2$ are obtained for the spread of $d_{z^2}$-like and $s$-like
MLWFs, respectively, the corresponding values constitute 55.78 and 319.44 bohr$^2$
for FSWFs. This indicates that the MLWFs differ significantly from the trial functions.

The reason for the rather large spreads of the FSWFs can be found by comparing the FSWF centers
to the MLWFs centers. In case of MLWFs the centers of the $s$-like WFs are located between
the atoms, forming covalent bridge-like Wannier functions. Such Wannier functions are
hard to construct directly from the atom-centered trial orbitals. The FSWFs constructed from 
the $s$- and $d_{z^2}$-like trial orbitals are, in contrast, located on the atoms, which causes
a significantly larger spread~\cite{MazariARXIV2011}. 

\begin{figure}
\begin{center}
\centerline{\includegraphics[width=0.35\textwidth,angle=270]{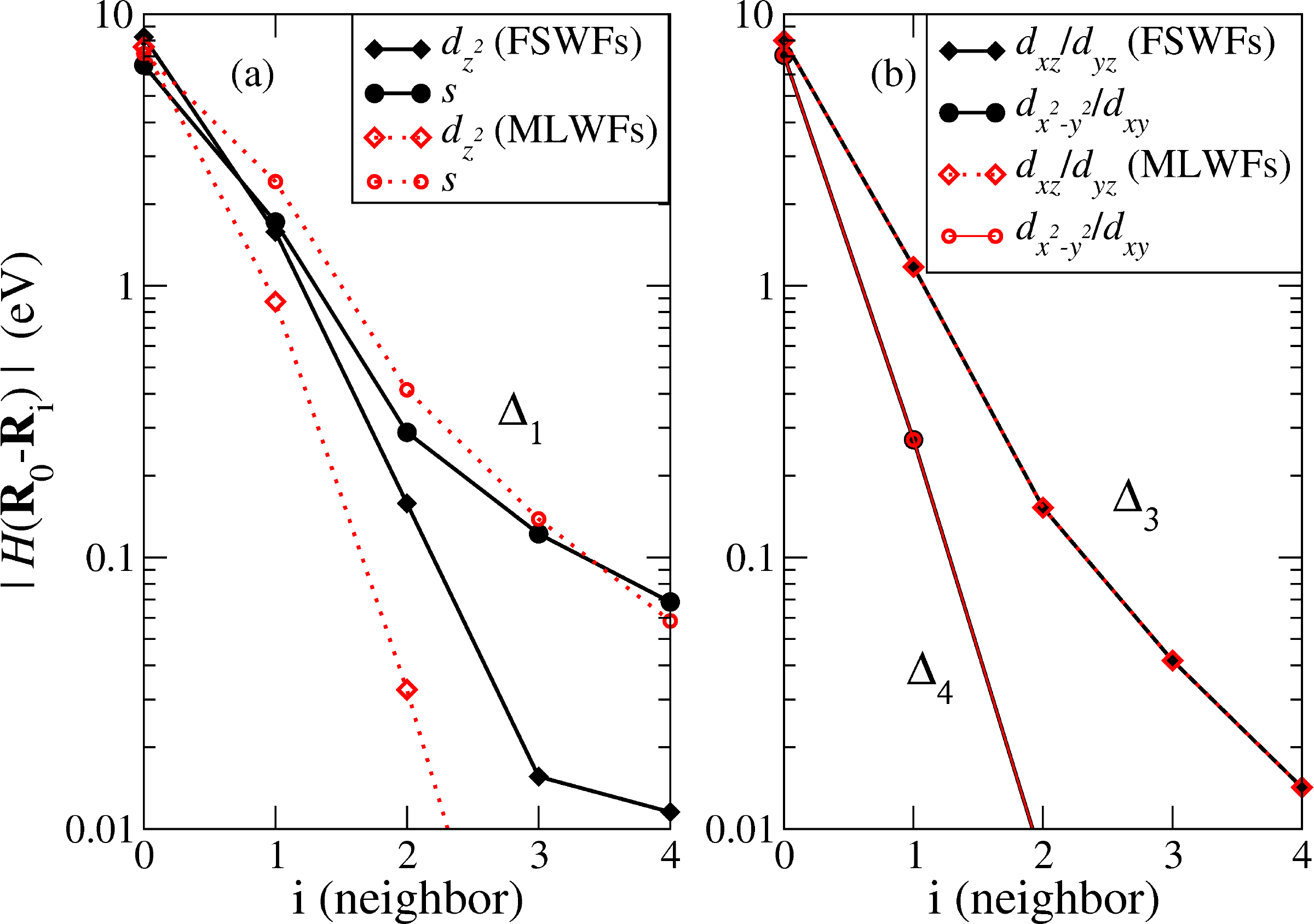}}
\caption{(color online)
Real space hopping integrals between orbitals
of the same type $|H(\mathbf R_i-\mathbf R_0)|$ as a function of the n.n. for a Pt monowire on a
logarithmic scale. The hoppings were calculated both with MLWFs (open red symbols)
and FSWFs (closed black symbols) for (a) $s$ and $d_{z^2}$ orbitals, and for (b)
$d_{xy}$, $d_{x^2-y^2}$, $d_{xz}$ and $d_{yz}$ orbitals.
}
\label{fig:MLWFs_vs_FSWFs_hop}
\end{center}
\vspace*{-1cm}
\end{figure}

In principle, all Hamiltonians obtained by mapping to Wannier functions
which include the hopping matrix elements between all WFs are equivalent. This equivalency 
is lifted, however, if we consider only a limited number of neighbors to set up our
tight-binding like Hamiltonian. In Fig.~\ref{fig:MLWFs_vs_FSWFs_BS},
the Pt monowire bandstructure based on the FLAPW-DFT calculation and
Slater-Koster interpolations of the bandstructure based on MLWFs and FSWFs are
compared. The trial orbitals for the FSWFs are in this case chosen to be
$s$- and $d$-like orbitals and centered on each atom. While in first n.n. approximation the
interpolated bandstructures differ between the MLWFs
and FSWFs approach, especially in the bandwidth of the more delocalized $s$ and
$d_{z^2}$ orbitals, already in second n.n. approximation both
WFs basis sets describe the FLAPW-DFT bandstructure equally well. 
By further
increasing the considered number of neighbors to third n.n. approximation,
the accuracy of the description increases
with respect to the $s$-bandwidth. However, the most important part with respect
to transport properties is the bandstructure in the vicinity of the Fermi level,
which does not improve significantly. For the more localized
$d_{xy}$, $d_{x^2-y^2}$, $d_{xz}$ and $d_{yz}$ orbitals even the first
n.n. description is sufficient as seen in the bandstructure and also from the
hopping matrix elements as seen in Fig. \ref{fig:MLWFs_vs_FSWFs_hop}.

At least for a rather simple system such as a perfect Pt monowire, the localization
procedure used to obtain MLWFs obviously does not influence the localized $d$-orbitals
mentioned above. Only the $s$ and $d_{z^2}$ states are affected, but the decay of
the hopping integrals is exponential irrespective of the description (FSWFs or MLWFs).
For systems more complicated than a Pt monowire, the initial
choice of trial orbitals may not be straightforward. In such a case the localization
procedure to obtain MLWFs significantly improves the accuracy of the calculation, while
for simpler systems where more intuitive choices of orbitals can be made FSWFs may be sufficient.
An example that both descriptions indeed lead  to very similar results 
with respect to transport calculations is
shown below for a Pt monowire with one elongated bond.
Note that the FSWFs
make the construction of the transport Hamiltonian, as discussed in
section~\ref{subsection:construction}, much more
simple, expecially for systems with a more complex electronic structure.

\subsection{Transmission: scalar-relativistic case}

\begin{figure}
\begin{center}
\centerline{\includegraphics[width=0.35\textwidth,angle=270]{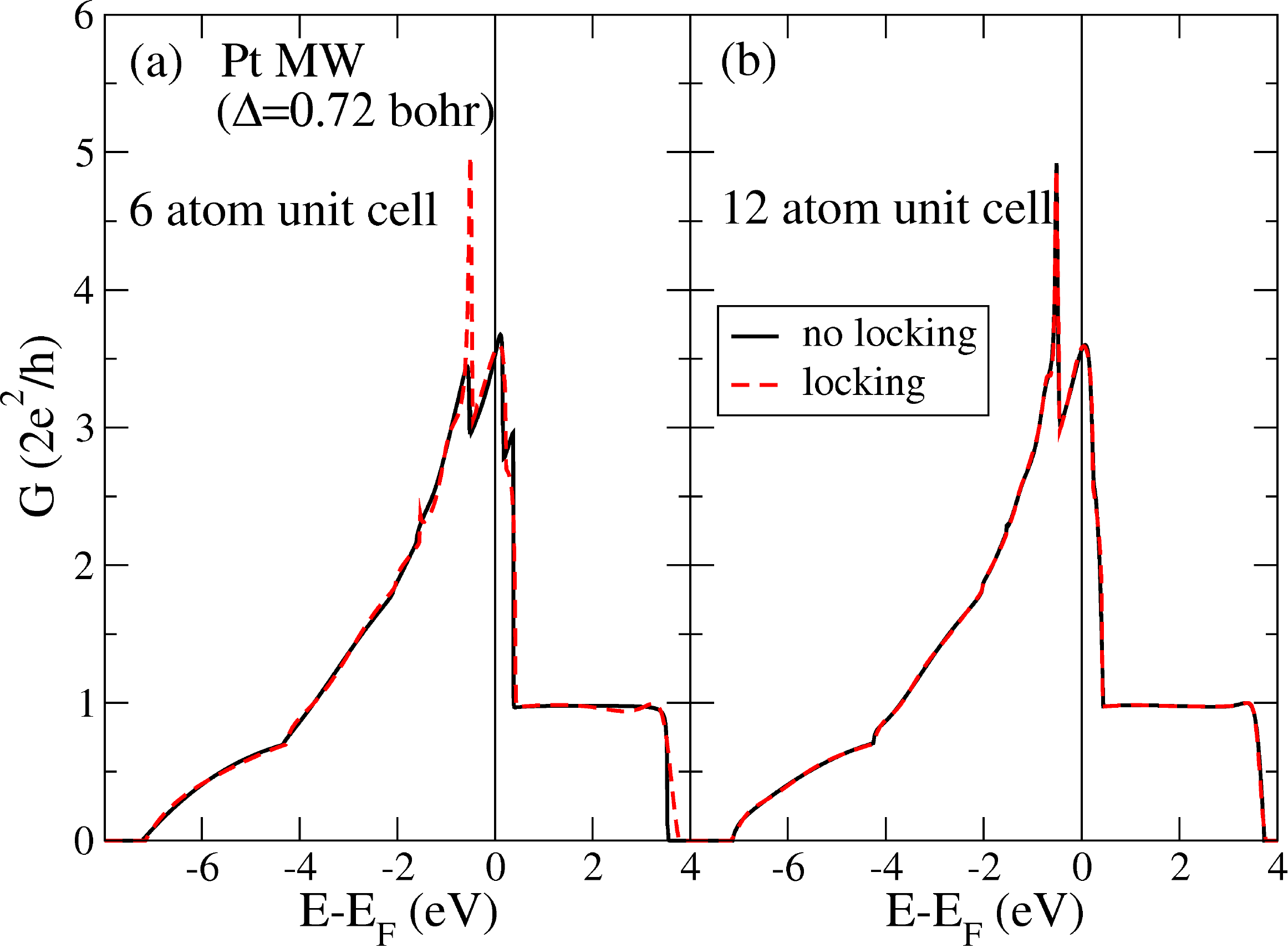}}
\caption{(color online)
Conductance for a Pt monowire with a single bond stretched by $\Delta=0.72$ bohr
using MLWFs within the nearest-neighbor approximation for the transport Hamiltonian and
(a) a 6 atom supercell and (b) a 12 atom supercell for the FLAPW calculation of the
scattering region. The semi-infinite leads have been described using the supercell
calculation (solid lines) or using the locking technique (dashed lines), i.e.,~using
perfect Pt monowires for the leads.
}
\label{fig:nl_vs_l}
\end{center}
\vspace*{-1cm}
\end{figure}

With the aid of the Pt monowire DFT calculations and the construction of WFs and the Hamiltonian from
the hopping matrix elements it is now possible to calculate the conductance based on the Green's function
method. We start by considering the quality of the locking technique. For this purpose, we compare the
results for a rather small 6 atom supercell calculation for the scattering region with
a single elongated bond of $\Delta= 0.72$ bohr and a calculation performed in a 12 atom supercell.
The quantum conductance obtained for both cases without applying the locking-technique, i.e.,~constructing the
semi-infinite leads from the supercell calculation, is similar but differs in key details such as a sharp
peak just below the Fermi energy (compare Fig.~\ref{fig:nl_vs_l}(a) and (b)).
If we replace the Hamiltonian for the leads by the one constructed from the MLWFs of a periodic Pt-monowire
the result changes as follows: While the conductance based on the 12 atom supercell calculation is nearly
independent on how the lead was constructed, the result for the 6 atom calculation improves significantly
upon using the locking technique and is almost indistinguishable from the calculation in the larger 12 atom supercell.
This demonstrates the applicability and quality of the locking technique which allows to save a considerable
amount of computational effort to calculate the ballistic transport properties.

\begin{figure}
\begin{center}
\centerline{\includegraphics[width=0.35\textwidth,angle=270]{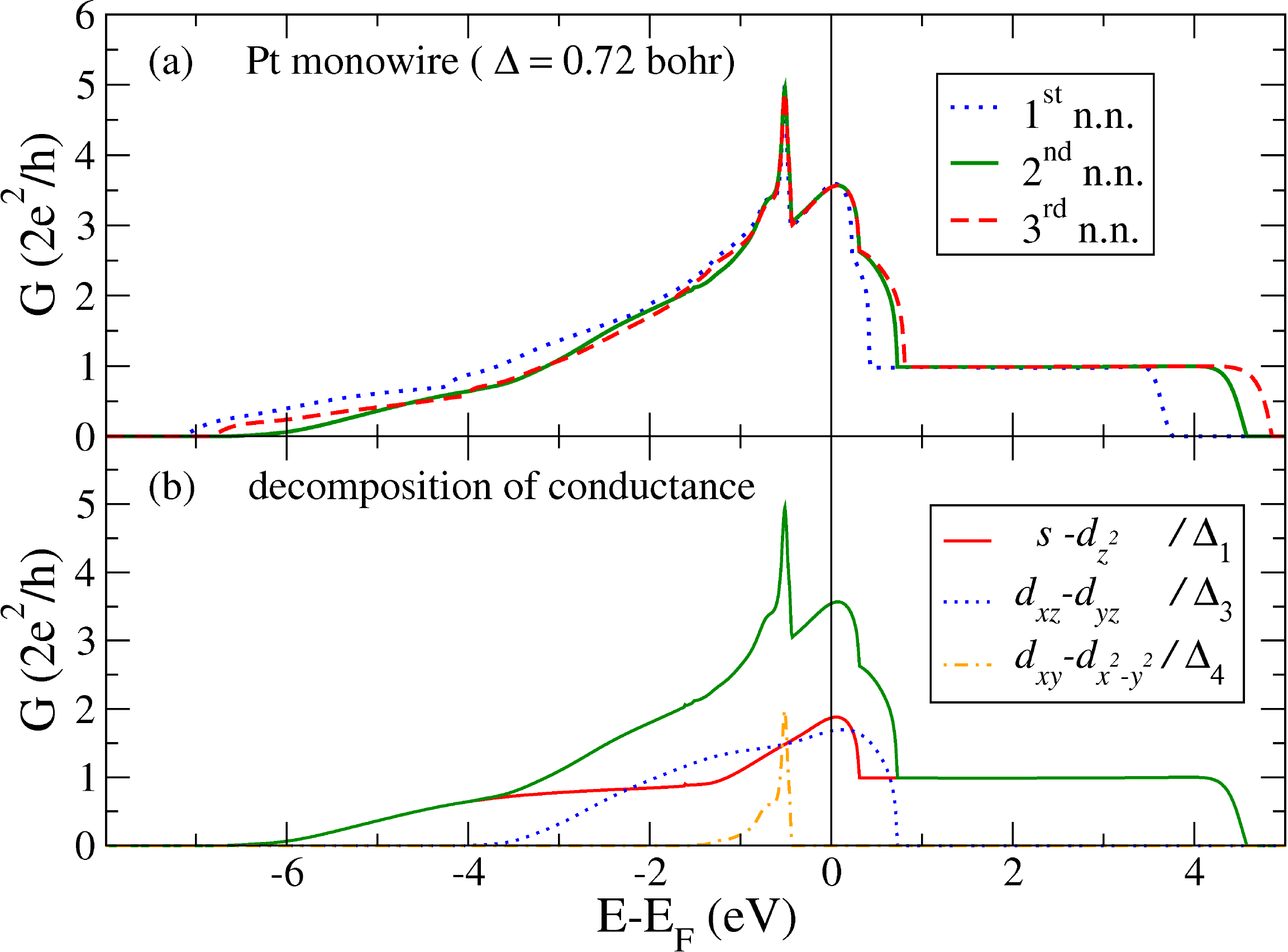}}
\caption{(color online)
(a) Conductance for a Pt monowire with a single bond stretched by $\Delta=0.72$
bohr based on a 12 atom supercell calculation using the first, second, and third
nearest-neighbor approximation for the construction of the transport Hamiltonian
from MLWFs and the locking technique to attach the leads. (b) Decomposition of
the total conductance (solid line) for the second nearest-neighbor approximation
into the contributions of the $s-d_{z^2}$ ($\Delta_1$) (solid red line), the
$d_{xz}-d_{yz}$ ($\Delta_3$) (dotted blue line) and the $d_{xy}-d_{x^2-y^2}$ ($\Delta_4$)
(dashed-dotted orange line) channels.}
\label{fig:1-to-3}
\end{center}
\vspace*{-1cm}
\end{figure}

While the previous test has been performed within the nearest-neighbor
approximation for the tight-binding like Hamiltonian we now determine how
accurate the calculated transmission function is with respect to the number
of neighbors included. In Fig. \ref{fig:1-to-3}(a) the transmission functions,
calculated in first, second, and third nearest-neighbor approximation and
based on the 12 atom supercell for the scattering region with one stretched
bond of $\Delta= 0.72$ bohr are presented.  The main effect which we observe
upon including more neighbors is a widening of the energy range in which the
conductance is non-zero as expected from the comparison of the bandstructure
obtained in the different approximations, cf.~Fig. \ref{fig:MLWFs_vs_FSWFs_BS}.
The conductance in the vicinity of the Fermi energy which is dominated by the
localized $d$-states is well described already using second nearest-neighbors.
Using only the first nearest-neighbor on the other hand results in an offset
of the conductance above the Fermi energy which originates from a shift of
the upper edge of the $\Delta_3$ band as seen in the bandstructure.  Therefore,
we use at least the second nearest-neighbor approximation in the following to
construct the tight-binding like Hamiltonian.

In order to understand which states contribute to the transmission we
can decompose it with respect to the orbital symmetry of the Wannier functions. The
individual transmission channels can be derived from Eqn.~(\ref{Transmission}), by
performing the trace operation only over WFs within the same symmetry group.  In
Fig.~\ref{fig:1-to-3}(b) we see that $s$-$d_{z^2}$ states provide an
almost perfectly conducting channel in a large energy range. Only far below the
Fermi energy the value drops below $2e^2/h$ and in the vicinity of the Fermi energy
it rises due to the availability of two $\Delta_1$ bands, cf.~the bandstructure
in Fig.~\ref{fig:MLWFs_vs_FSWFs_BS}.  The more localized $d_{xz}-d_{yz}$ states,
on the other hand, possess a much smaller transmission and their contribution
is localized in a small energy window. This effect is even more dramatic for the
$d_{xy}-d_{x^2-y^2}$ orbitals, which show a very small overlap and hopping matrix
elements leading to a sharp peak in the conductance.

\begin{figure}
\begin{center}
\centerline{\includegraphics[width=0.35\textwidth,angle=270]{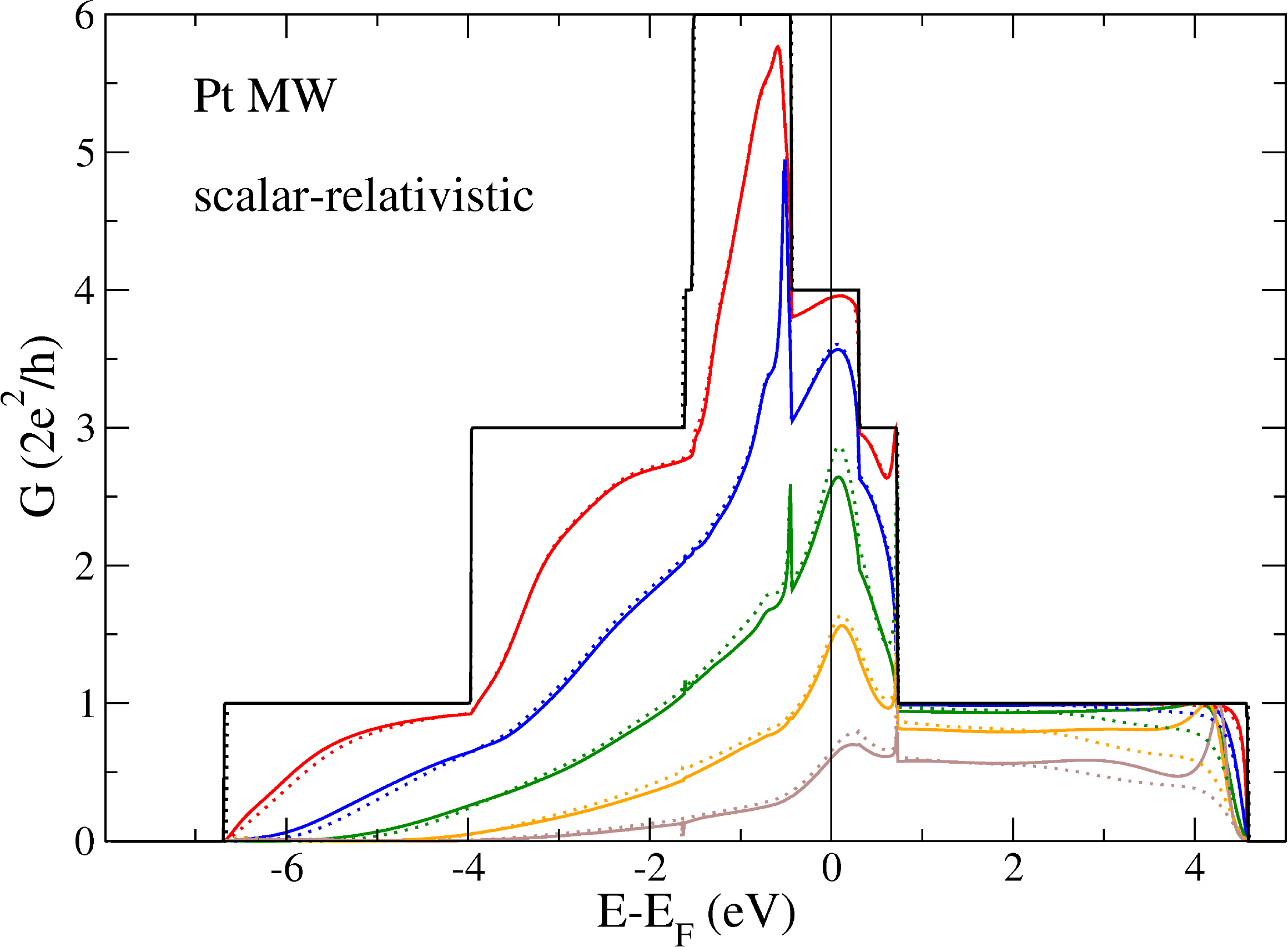}}
\caption{(color online)
Conductance for a nonmagnetic Pt monowire in the scalar-relativistic approximation, i.e.,~neglecting spin-orbit coupling,
with a single bond stretched by $\Delta$. The second nearest-neighbor approximation has been used for the transport
Hamiltonian. The WFs and hopping matrix elements have been constructed from a 12 atom supercell and the leads were
described by the locking technique. Curves are shown for MLWFs (solid lines) and FSWFs (dotted lines) for
$\Delta= 0.0$, $0.34$, $0.72$, $1.22$, $1.82$ and $2.52$ bohr (from left to right).}
\label{fig:fsT_vs_MLT}
\end{center}
\vspace*{-1cm}
\end{figure}

Finally, we turn to the conductance of the Pt monowire as a function of the
stretched bond length shown in Fig.~\ref{fig:fsT_vs_MLT}.
For the conductance of a perfect Pt wire, we find the expected step-function shape in which
each band contributes with one conductance quantum $G_0$ per spin within its bandwidth.
Upon increasing the length of a single bond in the wire, the overlap between the
Wannier orbitals across the gap decreases, especially for the more localized $d$-orbitals,
and as a result the transmission drops dramatically.  Accordingly, only the contribution
from the $s$-$d_{z^2}$ states survives at large gaps while the sharp peak originating
from the $d_{xy}-d_{x^2-y^2}$ orbitals vanishes above $\Delta=1.82$ bohr.
Another important result of this calculation is that the Hamiltonians obtained
with MLWFs and FSWFs provide nearly the same results, i.e.,~the radial solutions
of the FLAPW potential are evidently a reasonable choice as FSWFs trial orbitals.

\subsection{Transmission: spin-orbit coupling}
\label{subsec:Pt_SOC}

\begin{figure}
\begin{center}
\centerline{\includegraphics[width=0.40\textwidth,angle=270]{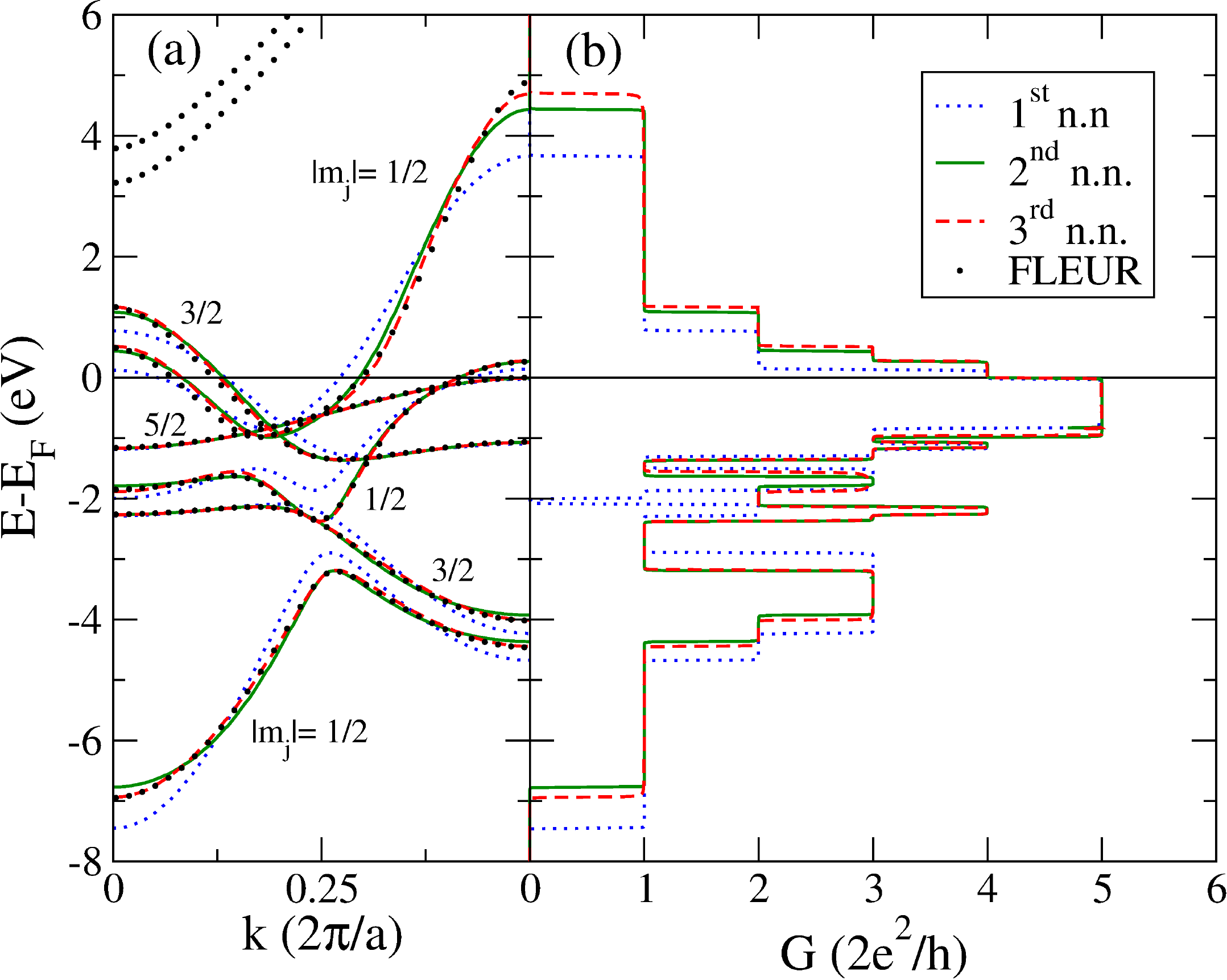}}
\caption{(color online)
Bandstructure of  an infinite nonmagnetic Pt monowire including spin-orbit coupling. (a)
 Bandstructure from the FLAPW calculation (big dots) and using the Hamiltonian from FSWFs within the
first, second, and third nearest-neighbor approximation.
(b) Conductance based on FSWFs for $1^{\mathrm{st}}$ (dotted line), $2^{\mathrm{nd}}$
(dashed line) and $3^{\mathrm{rd}}$ (solid line) n.n. approximation.
}
\label{fig:Pt_SOC_BS}
\end{center}
\vspace*{-1cm}
\end{figure}

For heavy transition metals such as Pt spin-orbit coupling plays an important role
and has a significant impact on the electronic structure. Evidently, the transport
properties should be equally affected. A suitable method to describe the quantum conductance
in such systems has to be capable of treating SOC. The effect of SOC on the electronic
structure, namely the coupling of the spin quantum number $s=\frac{1}{2}$ and
angular momentum quantum number $l=0,1,2,\dots$ to the total angular momentum quantum number
$j=\frac{1}{2}, \frac{3}{2},\frac{5}{2},\dots$ can be seen in Fig.~\ref{fig:Pt_SOC_BS}.
Compared to the scalar-relativistic calculation, in which SOC is neglected,
Fig.~\ref{fig:MLWFs_vs_FSWFs_BS}, the bandstructure including SOC changes significantly, 
Fig.~\ref{fig:Pt_SOC_BS}. In the chain geometry, the states are eigenfunctions
to the $z$-component (chain axis) of the total angular momentum and we can classify the
bands by the absolute value of $m_j$ as shown in Fig.~\ref{fig:Pt_SOC_BS} (a). Thereby,
spin-orbit coupling leads to several avoided crossings in the bandstructure,~e.g.~of
a $s-d_{z^2}$ and $d_{xz}/d_{yz}$-band around $3$~eV below the Fermi level.
With respect to the scalar-relativistic bandstructure, we also observe a significant
shift of the $d_{xy}$ and $d_{x^2-y^2}$-bands towards the Fermi energy. As this band
touches the Fermi energy at $k=\frac{\pi}{a}$ the conductance jumps from $4\,G_0$ in
the scalar-relativistic case to a value of $5\,G_0$. This finding already demonstrates
the importance of SOC for quantum transport calculations in such systems.

\begin{figure}
\begin{center}
\centerline{\includegraphics[width=0.35\textwidth,angle=270]{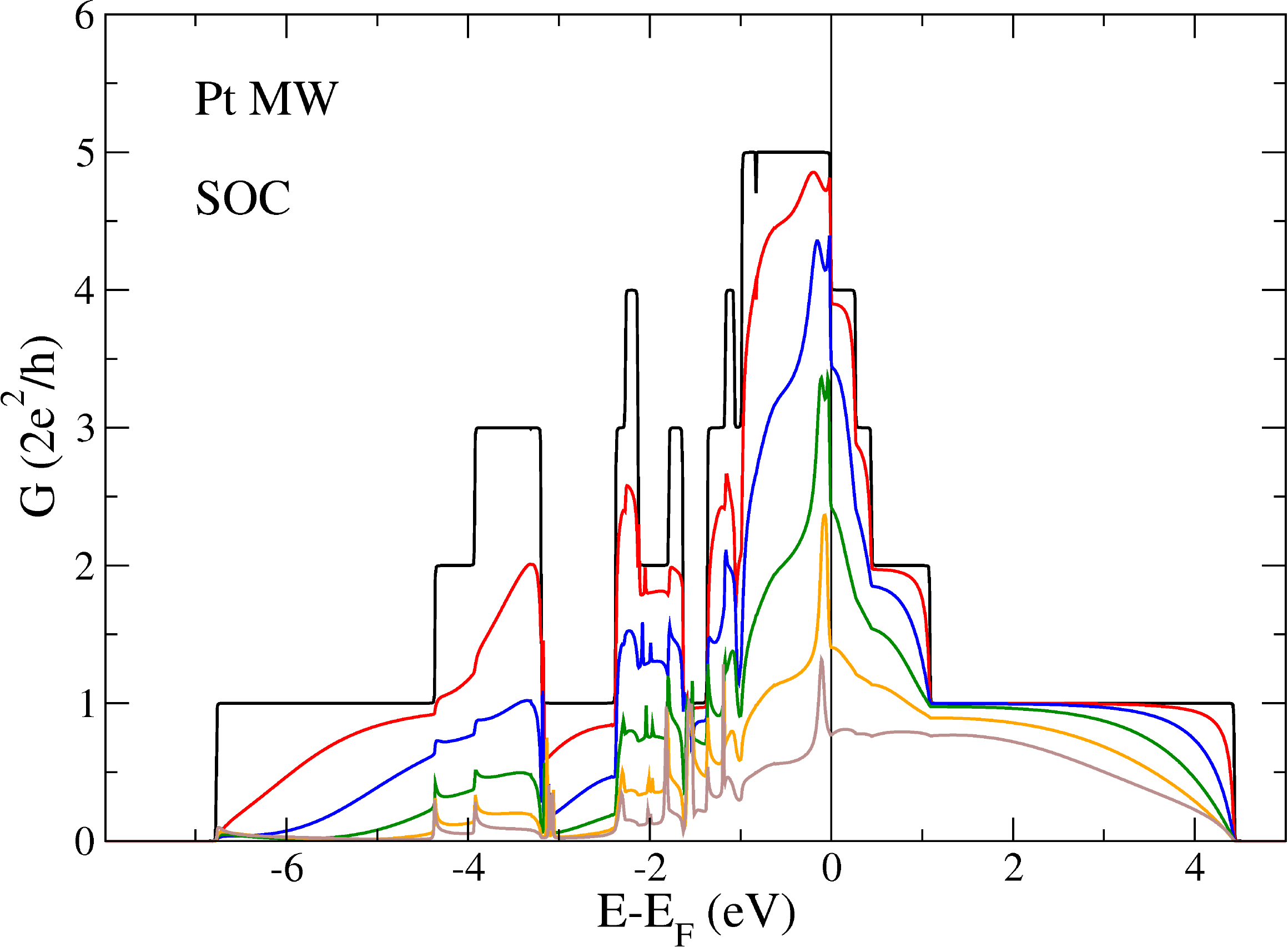}}
\caption{(color online)
Conductance for a nonmagnetic Pt wire with a single stretched bond $d_{\mathrm{Pt}}+\Delta$ including
spin-orbit coupling calculated within a 6 atom supercell and
using locking to semi-infinite Pt leads based on the Hamiltonian obtained from MLWFs
in second nearest-neighbor approximation.
From left to right: one bond stretched by $\Delta= 0.0$, $0.34$, $0.72$, $1.22$, $1.82$ and $2.52$ bohr.
}
\label{fig:Pt_SOC_T}
\end{center}
\vspace*{-1cm}
\end{figure}

The general form of the conductance in presence of SOC changes significantly, too, due to
the lifted degeneracies of bands with different $|m_j|$-values (see Fig.~\ref{fig:Pt_SOC_BS} (b)).
While the conductance at the Fermi level is enhanced upon taking SOC
into account, the degeneracy of the $d_{xy}$ and $d_{x^2-y^2}$-bands in the SR case leads
to a higher conductance of $6\,G_0$ below the Fermi energy.  Another key difference due to
SOC is the larger number of steps which appear in the conductance as a result of the anti-crossings
in the bandstructure, in particular, in the energy range of $3$~eV to $1$~eV below the Fermi level.
In Fig.~\ref{fig:Pt_SOC_T}, we also display the evolution of the conductance upon stretching a
single bond in the Pt monowire. Similar to the SR case, we observe a rapid decrease of the
conductance due to more localized $d$-orbitals. However, due to the spin-orbit split bands
there is a more pronounced peak structure in the conductance. In particular, we find a sharp
peak just below the Fermi energy which decays more slowly than in the SR calculation where it
is located slightly lower in energy.  Our calculations of the conductance are in good agreement
with those obtained based on fully relativistic ultrasoft pseudopotentials
and a scattering approach to obtain the conductance~\cite{DalCorsoPRB2006}.

\section{$\rm{Co}$ monowires}
\label{sec:Co_CD}

Another important aspect in transport through nanoscale structures is the effect of
spin-polarization and magnetic order.  Due to the reduced coordination number in nanostructures
the density of states is enhanced and according to the Stoner model the tendency towards
magnetism increases. The reduced symmetry also results in a much larger magneto-crystalline anisotropy
energy (MAE) as the orbital moments become more significant. E.g.~freestanding and suspended
chains of $4d$- and $5d$-transition-metals become magnetic and show giant values of 
the MAE~\cite{MokrousovPRL2006,Mokrousov2010}
and the effect of colossal magnetic anisotropy has been reported~\cite{SmogunovNN2008}.
Here, we demonstrate that our method allows spin-polarized transport calculations.
We consider a simple model system, i.e.,~a Co monowire with a single stretched bond and allow
a parallel and antiparallel alignment of the magnetization on the two
Co electrodes. We calculate the conductance in both configurations
and determine the magnetoresistance as a function of electrode separation.
The calculations in the antiferromagnetic configuration of the Co monowire can
also be compared to calculations by Smogunov {\sl et al.} based on the scattering approach and
pseudopotentials~\cite{SmogunovPRB2004}.

\subsection{Magnetoresistance}
 \label{subsec:Co_P_vs_AP}

\begin{figure}
\begin{center}
\centerline{\includegraphics[width=0.40\textwidth,angle=270]{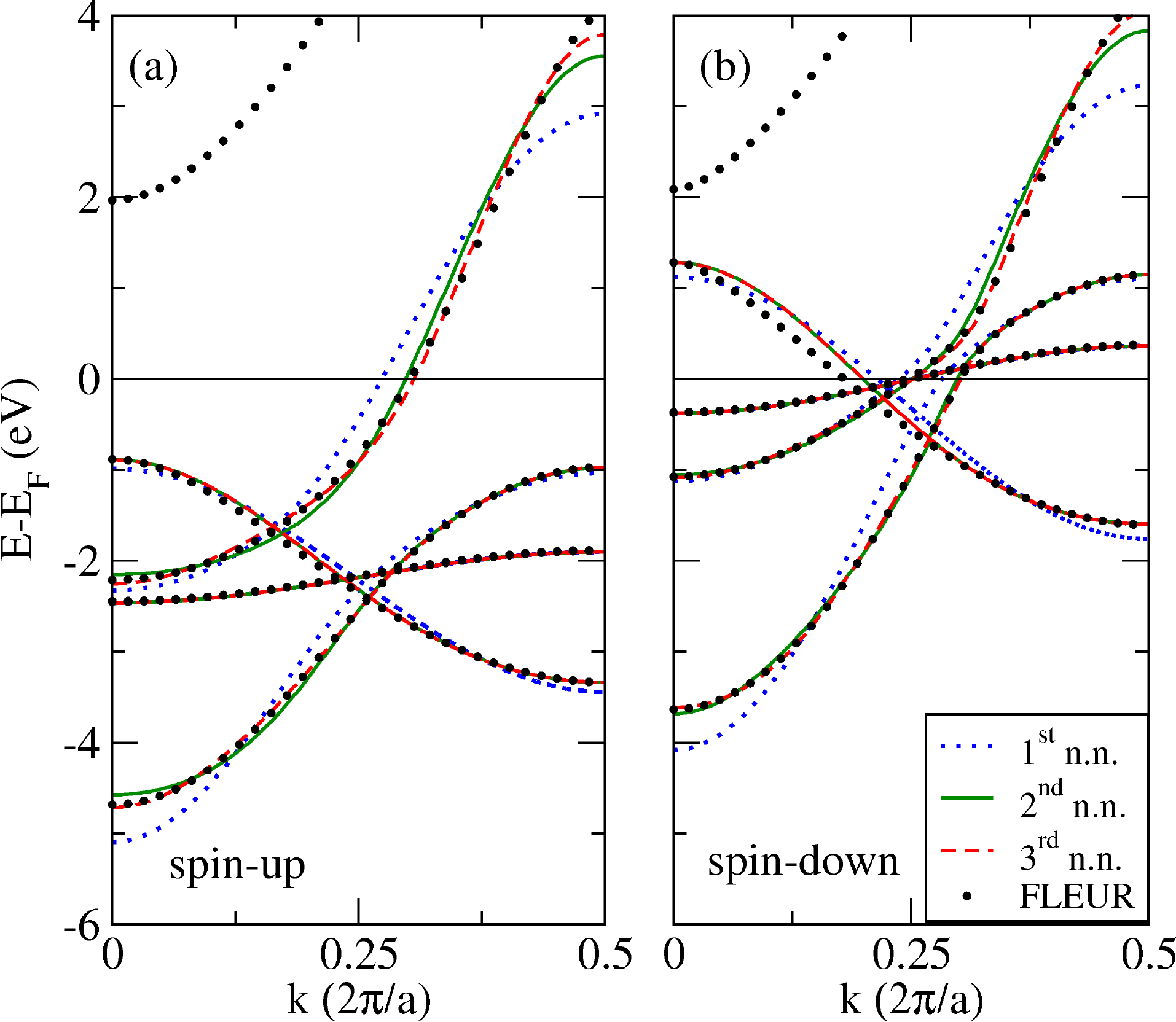}}
\caption{(color online)
(a) Majority and (b) minority bandstructure for a ferromagnetic Co monowire with
$d_{\rm Co}=4.15$ bohr calculated within FLAPW (big dots) and FSWFs in $1^{\mathrm{st}}$
(dotted lines), $2^{\mathrm{nd}}$ (dashed lines) and $3^{\mathrm{rd}}$ n.n. (solid lines)
approximation.
}
\label{fig:Co_DOS}
\end{center}
\vspace*{-1cm}
\end{figure}

\begin{figure}
\begin{center}
\centerline{\includegraphics[width=0.45\textwidth]{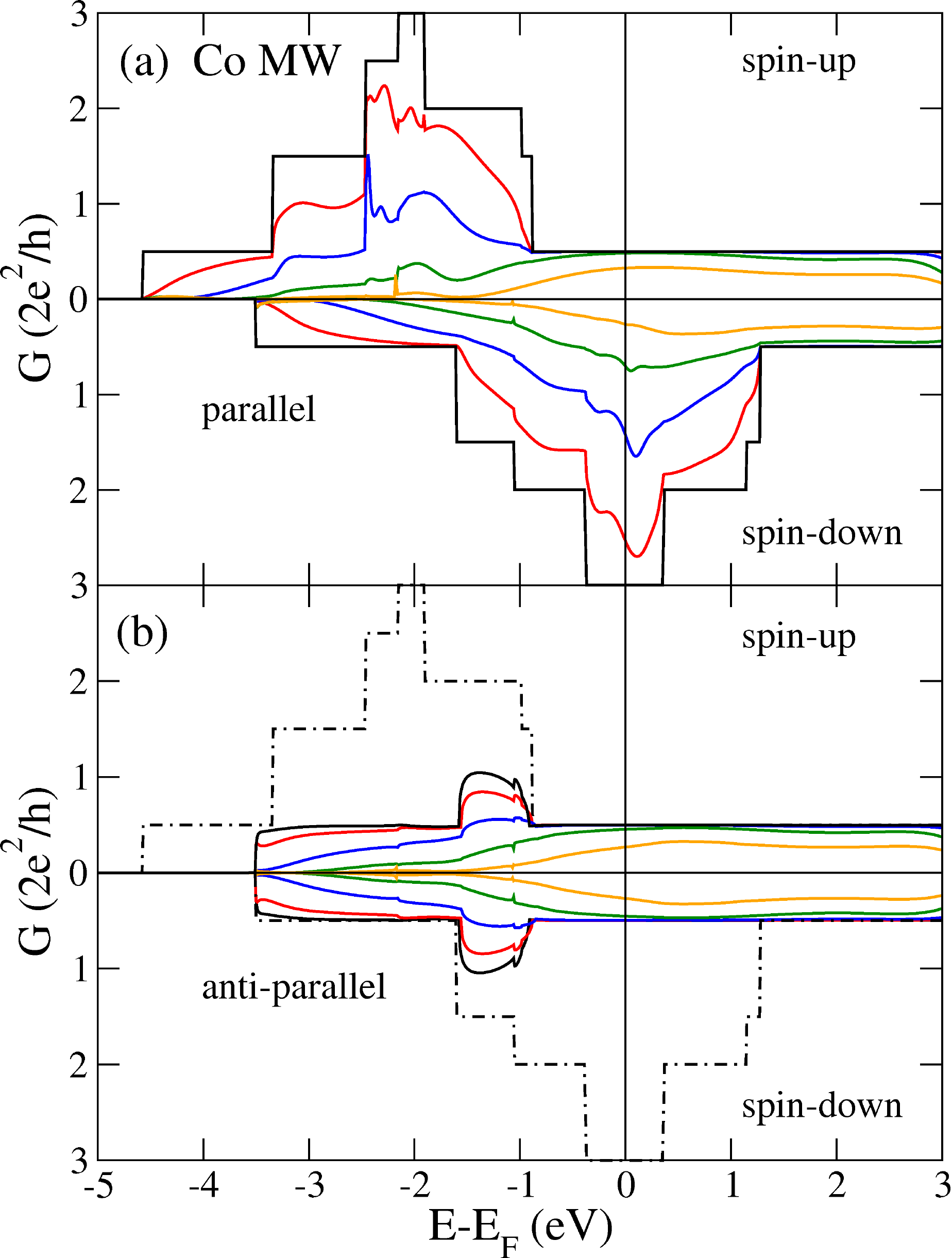}}
\caption{(color online)
Conductance between two ferromagnetic Co monowires separated by a gap, $\Delta$, in (a)
parallel and in (b) antiparallel alignment of the magnetization. A supercell of 16
atoms has been used for the scattering region and the transport Hamiltonian was constructed
based on FSWFs in the second n.n. approximation. From left to right: gap of $\Delta= 0.0$,
$0.45$, $1.05$, $1.85$ and $2.85$ bohr.  Upper and lower part of the plots show the spin-up
and spin-down transmission channel, respectively.
}
\label{fig:Co_P}
\end{center}
\vspace*{-1cm}
\end{figure}

Compared to the non-magnetic Pt bandstructure, the Co chain exhibits a smaller band width 
due to more localized $3d$-states, and a large exchange splitting (Fig.~\ref{fig:Co_DOS}).
The exchange splitting leads to a net spin moment in the unit cell of 2.13$\mu_B$.  
A good overall accuracy in reproducing this bandstructure based on FSWFs can be achieved
if we go up to third nearest-neighbor hoppings. For the $d$-bands 
and the $s$-$d_{z^2}$ bands around the Fermi
energy even the second nearest-neighbor approximation is sufficient. From the spin-split
bandstructure we expect a larger conductance in the parallel
magnetization alignment due to the overlap between minority bands
of $\Delta_3$ and $\Delta_4$ symmetry. This notion is confirmed
by the calculated conductance in the two magnetic configurations
as a function of gap size as shown in Fig.~\ref{fig:Co_P}.
At the Fermi level, we observe majority and minority spin
conductances of $G_{\rm maj}=e^2/h$ and $G_{\rm min}=6 e^2/h$, respectively, for
a perfect ferromagnetic Co monowire (see Fig~\ref{fig:Co_P} (a)). As the central bond is
stretched the minority conductance drops rapidly because it originates from the
more localized $d_{xz,yz}$- and $d_{xy, x^2-y^2}$-states. The majority
conductance, on the other hand, is due to $s-d_{z^2}$-states and decays much more slowly.

In the antiparallel alignment (Fig~\ref{fig:Co_P} (b)), the conductance is the same
in both spin channels. There is only a small energy
window between $1$~eV and $1.7$~eV below the Fermi energy
in which the $d_{xz,yz}$- and $d_{xy, x^2-y^2}$-states overlap and at the Fermi
energy, the conductance is dominated by the
$s-d_{z^2}$-states. The conductance in the antiparallel alignment
can be interpreted as an
envelope of spin-up and spin-down transmission functions calculated
for the parallel case as an electron can only be transmitted if there
are states of the same symmetry in both spin channels. The conductance
in this configuration is also in good agreement with that reported
by Smogunov {\sl et al.} using a scattering
approach and pseudopotentials~\cite{SmogunovPRB2004}.

\begin{figure}
\begin{center}
\centerline{\includegraphics[width=0.35\textwidth,angle=270]{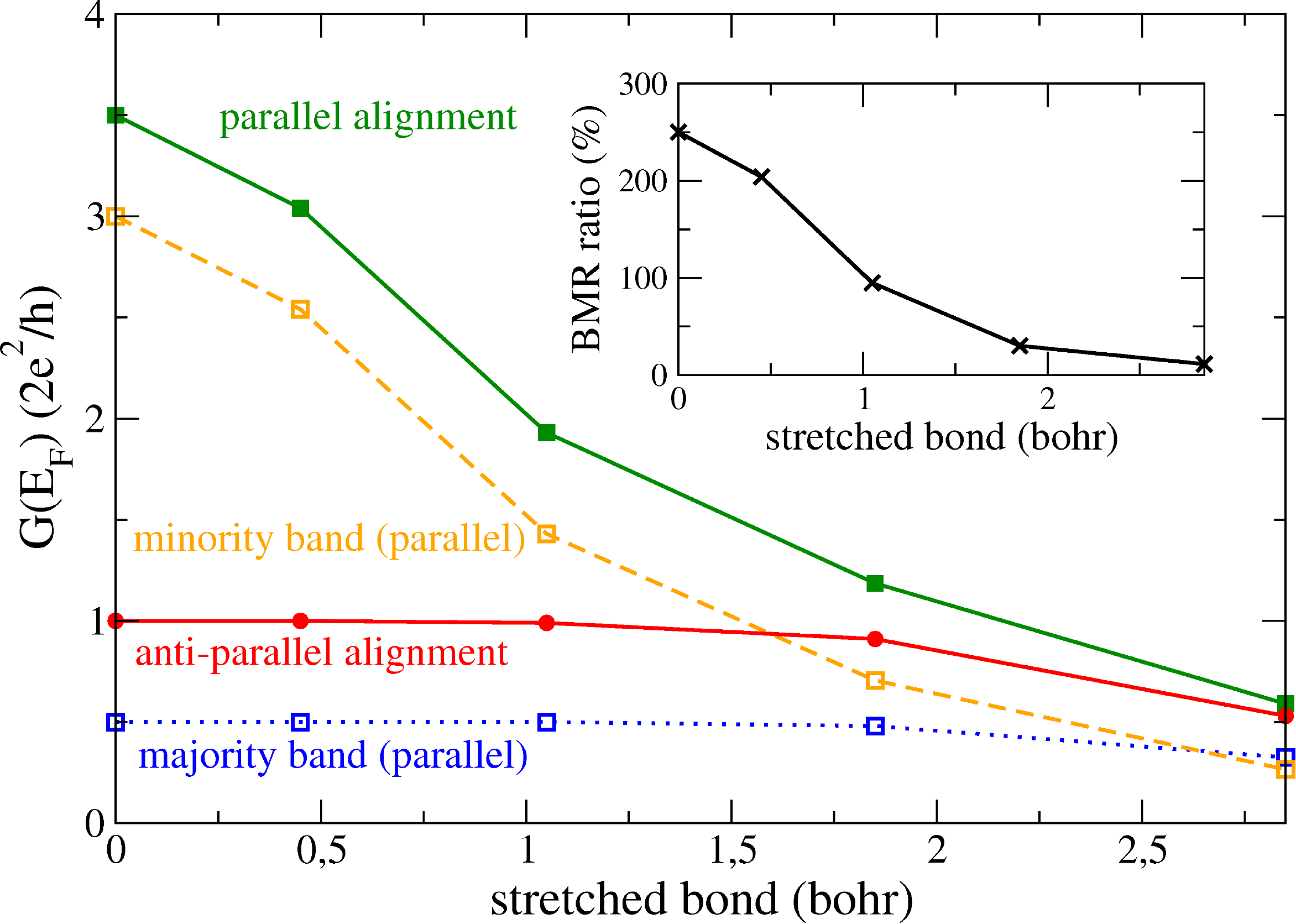}}
\caption{(color online)
Conductance at $E_F$ for the antiparallel (red filled circles, solid
line) and the parallel (green solid squares, solid line)
alignment of two ferromagnetic Co monowires as a function of
separation. For the parallel case the decomposition into majority
(blue open squares, dashed line) and minority (orange open squares,
dotted line) contributions is given.  The inset shows the BMR ratio
as a function of separation. 
}
\label{fig:Co_TMR}
\end{center}
\vspace*{-1cm}
\end{figure}

Based on the obtained quantum conductance at the Fermi level we
can calculate the ballistic magnetoresistance (BMR) upon stretching
the central bond.  The BMR is defined as the difference between the
conductance in the parallel and antiparallel alignment divided
by the antiparallel conductance:
\begin{equation}
\mathrm{BMR}=\frac{G_\mathrm{P}(E_F)-G_{\mathrm{AP}}(E_F)}{G_{\mathrm{AP}}(E_F)}\times 100\%
\end{equation}
Fig.~\ref{fig:Co_TMR} displays the evolution of the
spin-resolved conductance as a function of gap size for the two magnetic
configurations. As noted above, the parallel alignment is
characterized by a rapidly decreasing minority spin conductance and
a nearly constant majority spin contribution. However,
the minority spin conductance dominates until the end of the bond length
range which we considered. In the antiparallel alignment, the
conductance of both spin channels is the same and behaves similar
to the majority spin channel of the parallel alignment as it
is due to $s-d_{z^2}$-states. From this analysis of the channel
contribution we can understand the fast drop of the BMR
found upon stretching (inset of Fig.~\ref{fig:Co_TMR})
of the central bond in the monowire.

\section{Spin-orbit scattering at impurities}
\label{subsec:SOC_in_MJs}

In the previous sections we applied our quantum transport code to systems
with strong spin-orbit coupling (Pt monowires) and high spin-polarization
(Co monowires). In the following we combine the two effects in order to study the scattering
at impurities in the presence of spin-orbit coupling. We consider
two types of model systems. We begin with non-magnetic
Pt monowire with a single Co impurity atom and calculate the
dependence of the conductance on the magnetization direction
of the Co atom. An analysis of the orbital decomposed transmission
function allows us to study the influence of SOC on the
different channels. We find that band mixing due to SOC
has a pronounced influence, in particular, on the contribution
from the band with $\Delta_4$-symmetry. As a second system,
we consider a ferromagnetic Co monowire with a single Pt impurity
atom and compute the conductance for the two magnetization
directions of the Co wire, either along the direction of the wire
or perpendicular to it. From our calculations of the conductance
including spin-orbit coupling we can also determine
the ballistic anisotropic magnetoresistance (BAMR),
i.e.,~the difference of transmission between a magnetization parallel
to the current and perpendicular to the current.

While our systems are idealized they can be seen as prototypical for
experiments which may be performed for example by scanning tunneling
microscopy in the contact regime~\cite{ZieglerNJP2011}
or in break junctions~\cite{SokolovNN2007,KizukaPRB2008,Mokrousov2010}.
Scalar-relativistic calculations, i.e.,~neglecting SOC, in a similar geometry for a Ni
impurity in a Au monowire have been performed before\cite{MiuraPRB2008}.

\subsection{Magnetic impurity in a non-magnetic wire}
\label{subsec:Pt-Co-Pt}

We begin our investigation of spin-orbit scattering at
an impurity by considering a single Co atom in a Pt monowire. This
is the simpler of the two systems due to the nonmagnetic
Pt leads. We have already
discussed the conductance of Pt monowires with and without
spin-orbit coupling in section~\ref{sec:Results}.
Here, we study the conductance for different magnetization
directions of the Co impurity atom in order to calculate the so-called
ballistic anisotropic magnetoresistance (BAMR), which has been
predicted based on DFT calculations~\cite{VelevPRL2005} and was
experimentally reported for Co break junctions~\cite{SokolovNN2007}.

\begin{figure*}
\begin{center}
\centerline{\includegraphics[width=0.74\textwidth,angle=270]{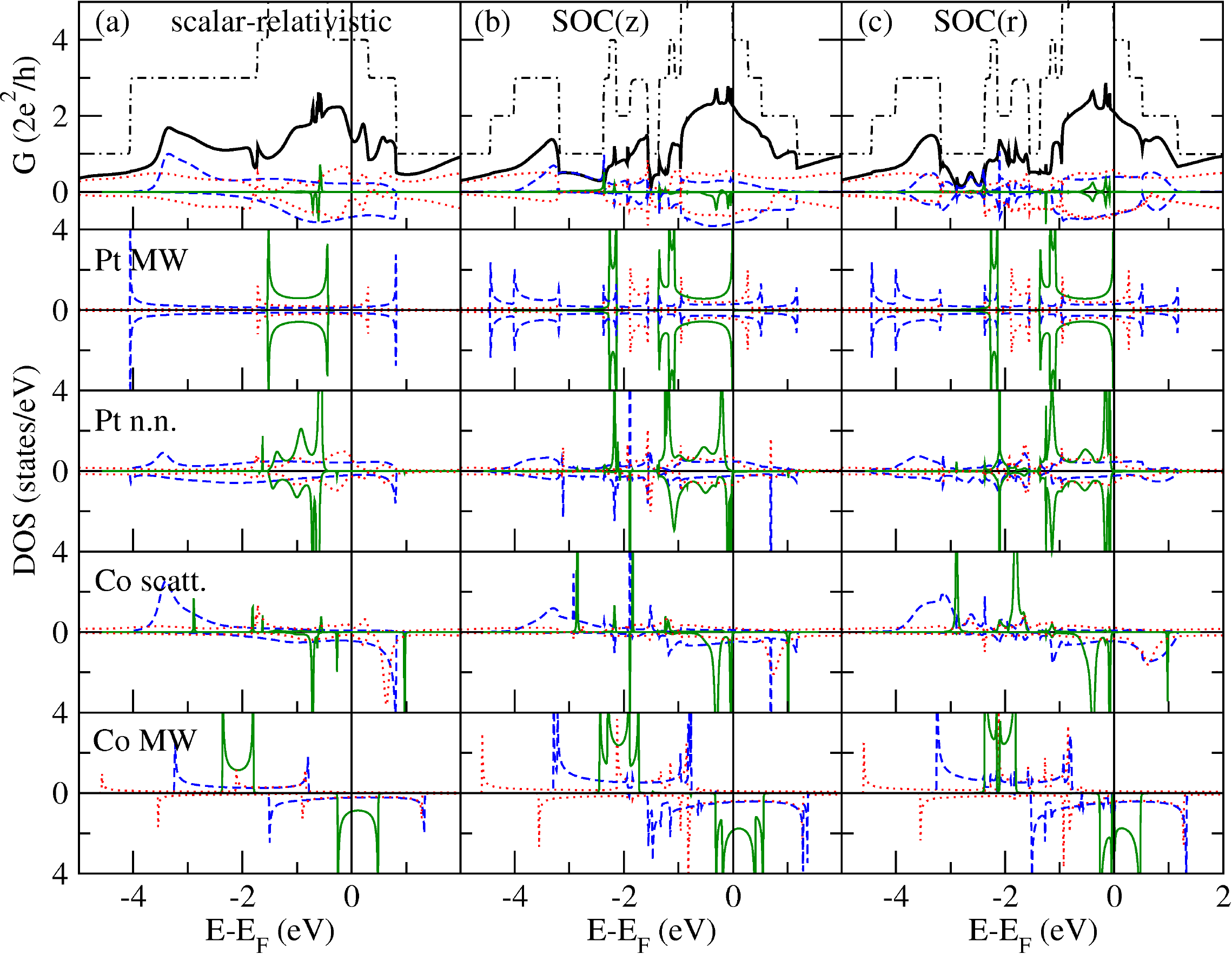}}
\caption{(color online)
Conductance of a Pt monowire with a single Co impurity in (a) the
scalar-relativistic (SR) approximation and including spin-orbit
coupling (SOC) for a magnetization (b) along the chain axis ($z$) and
(c) perpendicular to it ($r$).  In addition to the total conductance
(black thick line) each panel shows the transmission for a perfect Pt monowire
(dashed-dotted line) and orbital-decomposed into the $\Delta_1$-
(red dotted line), $\Delta_3$- (blue dashed line), and $\Delta_4$-band
(green solid line) contribution. The projection onto the spin-up
and spin-down states is given for two different directions of the
$y$-axis, respectively. Below each conductance panel
the density of states (DOS) is displayed
in the corresponding electronic configuration, i.e.,~SR or SOC,
for a perfect Pt monowire, the Pt atom adjacent to the Co
impurity, the Co impurity, and a perfect Co monowire. The DOS
is orbital-decomposed similarly to the transmission.
}
\label{fig:Pt-Co-Pt_G_vs_DOS_incl_SOC}
\end{center}
\vspace*{-1cm}
\end{figure*}

Before we discuss the calculated conductance, we focus
on the magnetic properties of our system.
From the DFT calculations in the scalar-relativistic
case we obtain spin moments of 2.46$\mu_B$ for the Co atom which induces
Pt spin moment of a magnitude
of up to 0.27$\mu_B$, oscillating in sign 
as a function of separation from the Co atom.
A similar behavior 
was found upon including SOC in the calculations for both magnetization
directions, with a Co spin moment of about 2.49$\mu_B$.
Including spin-orbit interaction in the calculations gives rise to finite values 
of the orbital moments of the atoms, which play an important role in determining the
energetically favorable direction of the magnetization.\cite{MokrousovPRL2006}
In our system, the orbital moments of the Co atoms are much larger than those
of the surrounding Pt atoms, and constitute 0.12$\mu_B$ and 0.19$\mu_B$ for the 
magnetization along the chain axis ($z$) and perpendicular to it ($r$), respectively. 
Accordingly,\cite{MokrousovPRL2006}
this results in an energetical preference of the in-chain magnetization direction
over the out-of-chain direction, with a calculated magneto-crystalline anisotropy
energy (MAE) of $4.3$~meV per magnetic atom.

We now turn to the calculated conductance presented in the
three top panels of Fig.~\ref{fig:Pt-Co-Pt_G_vs_DOS_incl_SOC} 
for the scalar-relativistic case
and upon including spin-orbit coupling for the two different magnetization directions.
For reference the orbitally decomposed conductance and the density of states (DOS)
of a perfect Pt monowire is given in each of the three plots and in the panels below, respectively.
As a general trend, the introduction of a Co scatterer results	in a
non-perfect matching between the spin-split Co $3d$-states and
the more delocalized Pt $5d$-states (cf.~the bandstructures in
Figs.~\ref{fig:MLWFs_vs_FSWFs_BS} and \ref{fig:Co_DOS}).
In all three cases, a clear signature of the exchange-split
Co $\Delta_3$-band can be observed in the overall conductance,
most clearly visible in the spin- and orbital
decomposition. As expected, the $\Delta_4$-bands are 
shifted towards the Fermi energy upon including spin-orbit coupling, 
but due to the energetical mismatch between the Co and Pt $\Delta_4$-bands in 
SR and for both magnetization directions with SOC, this band plays only a minor role in the overall conductance. 

\begin{figure}
\begin{center}
\centerline{\includegraphics[width=0.35\textwidth,angle=270]{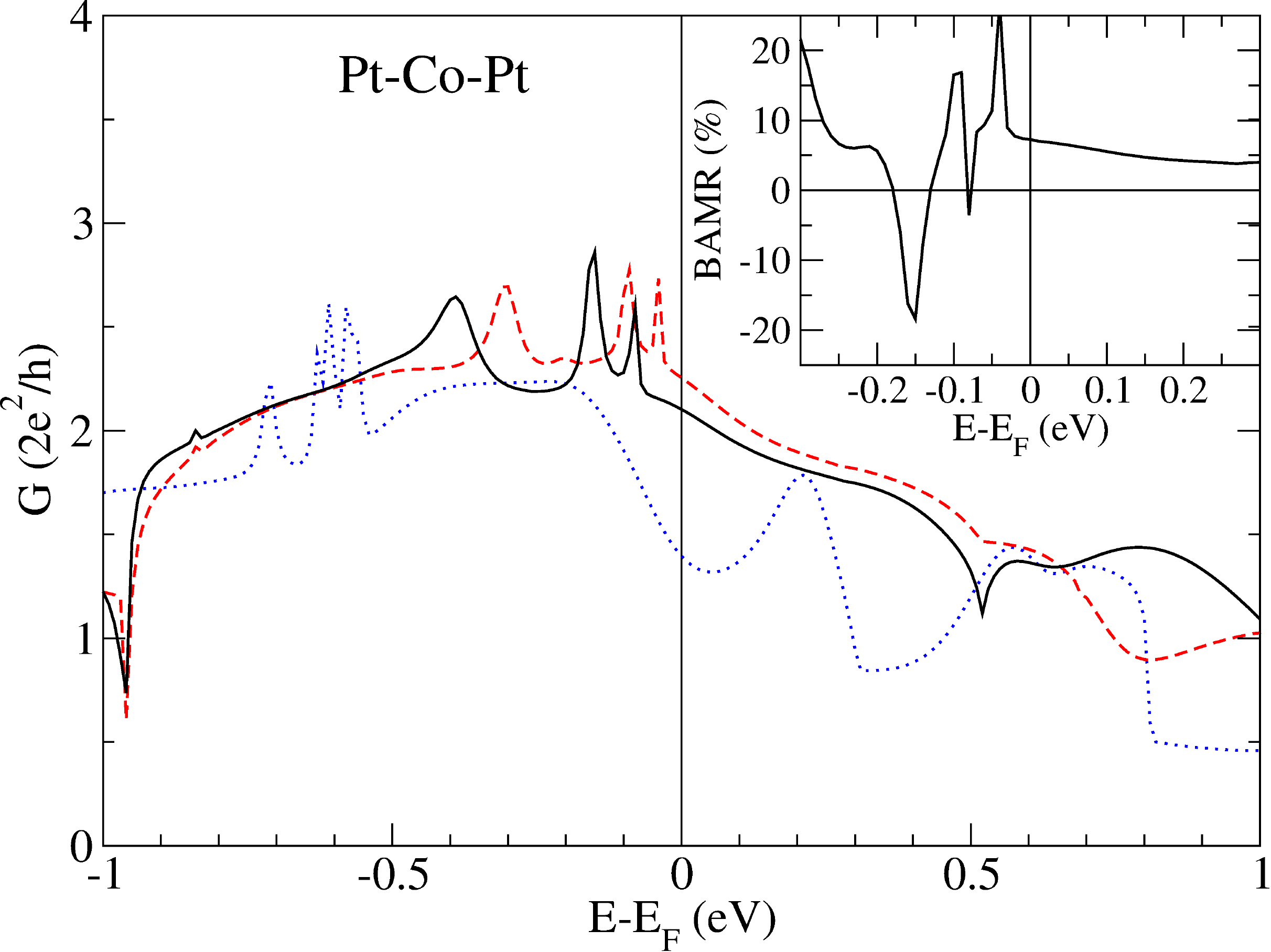}}
\caption{(color online)
Conductance around the Fermi level for a Pt wire with a single Co impurity atom without spin-orbit coupling (dotted blue line) and
including SOC for an in-chain (dashed red line) and an out-of-chain (solid black line) magnetization direction.
The inset shows the ballistic anisotropic magnetoresistance
(BAMR) as defined by Eqn.~\ref{eqn_BAMR}.
}
\label{fig:BAMR}
\end{center}
\vspace*{-1cm}
\end{figure}

Nevertheless, there is a considerable difference between the
conductance at the Fermi level in the scalar-relativistic case,
$G_{SR}=1.40\,G_0$, and upon including SOC either for  $z$-magnetization,
$G_{\parallel}=2.25\, G_0$, or $r$-magnetization, $G_{\perp}=2.10\, G_0$, as seen 
in Fig.~\ref{fig:BAMR}.
The main reason for this large difference between SR and SOC conductances can be found in the $\Delta_1$ band of 
SR Pt. In this channel the DOS is reduced compared to the SOC cases at the Fermi energy at the Pt n.n. atoms and
there is a corresponding reduction of the conductance, as shown in 
Fig.~\ref{fig:Pt-Co-Pt_G_vs_DOS_incl_SOC}.
The difference of $G_{\parallel}-G_{\perp}=0.15\, G_0$ between the two different magnetization directions can be found
in the larger minority $\Delta_3$-state contribution of the parallel aligned axis. Here, the SR and the 
parallel SOC case behave similarly. The DOS for $\Delta_3$ majority states is small at the Fermi level,
the majority state conductance is reduced in comparison to the minority state contribution, as a result
of the exchange splitting of the Co scatterer.

Interestingly this is not the case for the $r$ direction of the magnetization, for which  
majority and minority channels contribute equally to the  total conductance. 
This effect also occurs for the
$\Delta_3$ minority channel between $-2.8$~eV and $-3.9$~eV as well as for the $\Delta_4$ conductance just  below
the Fermi energy. While the very sharp spin-up $\Delta_4$-peak in the SR transmission at $-$0.7~eV
can be traced back to a small spin-up $\Delta_4$ peak in the DOS of the central Co 
atom at this energy, this is not the case for the mentioned regions in case of the $r$-magnetization,
for which no majority $\Delta_3$ and $\Delta_4$ states are present at  
the scatterer. 
The origin of this effect is the broken cylindrical symmetry when the magnetization points out of chain. 
This broken symmetry allows for a hybridization between $\Delta_1$ and $\Delta_3$ bands with
$j=\frac{1}{2}$, as well as between $\Delta_3$ and $\Delta_4$ bands with $j=\frac{3}{2}$. As a result an incident electron of
$j=\frac{1}{2}$ ($\frac{3}{2}$) can be transmitted into a state with $j=\frac{1}{2}$ ($\frac{3}{2}$) of different orbital character and spin.
This channel for scattering is less effective than the spin-conserving scattering for the in-chain magnetization,
resulting in a larger conductance in this case.

The changes in the ballistic conductance due to ballistic spin-scattering are important for the ballistic anisotropic magnetoresistance (BAMR).
The BAMR is defined analogously to the anisotropic magnetoresistance as:
\begin{equation}
\label{eqn_BAMR}
{\mathrm{BAMR}}=\frac{G_{\parallel}-G_{\perp}}{G_{\perp}}\times 100\%,
\end{equation}
where $G_{\parallel}$ and $G_{\perp}$ are the conductances for the magnetization
along the wire axis and perpendicular to it, respectively~\cite{VelevPRL2005}.
The difference of $0.15\, G_0$ at the Fermi level in favor of the parallel quantization axis
due to ballistic spin-scattering results in a small BAMR of the order of 7$\%$, see inset of
Fig.~\ref{fig:BAMR}. A small shift between the $\Delta_4$ contributions due to a small spin-splitting of those
bands for the Pt atom neighboring the Co scatterer (cf. Fig.~\ref{fig:Pt-Co-Pt_G_vs_DOS_incl_SOC}) results in an oscillatory behaviour of the BAMR when the
energy is varied from $-0.05$~eV to $-0.2$~eV, with BAMR ranging from $-$20\% to
25\%.

\subsection{Non-magnetic impurity in ferromagnetic wire}
\label{subsec:Co-Pt-Co}

\begin{figure*}
\begin{center}
\centerline{\includegraphics[width=0.64\textwidth,angle=270]{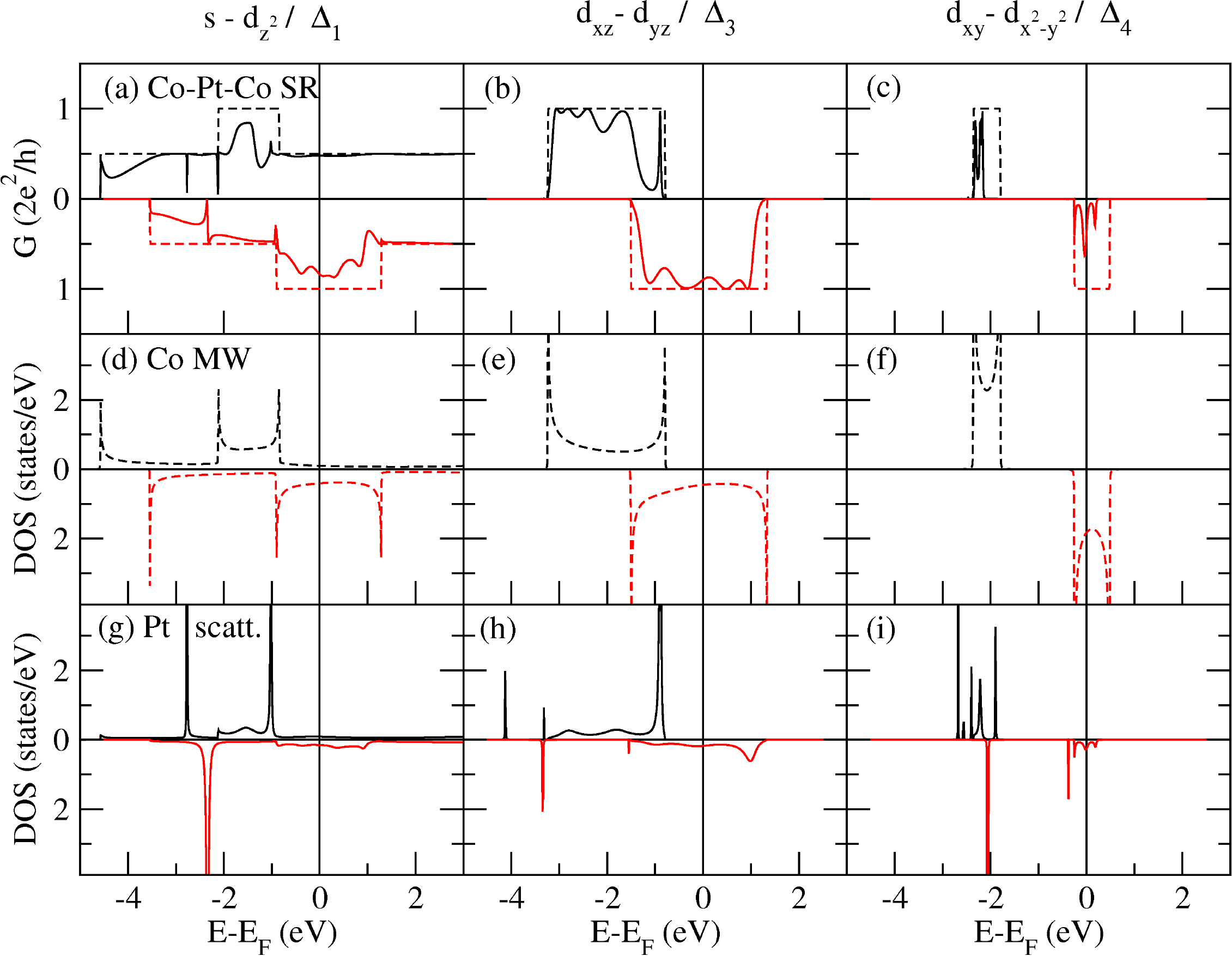}}
\caption{(color online)
(a-c) Orbital decomposition of transmission through a ferromagnetic Co wire with a single Pt impurity (solid lines) and a perfect Co wire
(dashed lines) for the $\Delta_1$, $\Delta_3$, and $\Delta_4$ channels, respectively, for spin-up (black, upper part) and spin-down
(red, lower part). (d-f) Density of states of the Co leads orbitally decomposed for spin-up (dashed black lines, upper part) and spin-down
(red dashed lines, lower part) states. (g-i) Orbitally decomposed DOS of the Pt impurity atom, for spin-up (black, upper part) and spin-down (red, lower part).
}
\label{fig:Co-Pt-Co_SR}
\end{center}
\vspace*{-1cm}
\end{figure*}

In the previous example, we have seen that the transmission 
can be affected by ballistic spin-scattering, leading to a small BAMR below the Fermi energy and BAMR 
oscillations due
to a shift in the $\Delta_4$ orbitals of the Pt atom next to the Co scatterer.
In this section, we consider a non-magnetic scatterer, a
Pt atom, in a ferromagnetic Co monowire. We find that this situation leads
to an enhanced BAMR close to the Fermi level, which is crossed by the $\Delta_4$-band.
In this case we do not expect strong ballistic spin-scattering because of the magnetic leads,
since large exchange splitting prohibits scattering between the states with opposite spin. 

First we consider the junction in the scalar-relativistic
approximation in order to understand the main impact of
the Pt scatterer on the conductance. While Co atoms in the
leads carry a magnetic moment of $2.13 \mu_B$, the Co atoms in the vicinity
of the Pt atom have moments in the range of 2.15$-$2.20$\mu_B$, and
the Pt atom itself is spin-polarized with a considerable moment of $0.36\mu_B$. As can be
seen in the orbitally decomposed conductance, Fig.~\ref{fig:Co-Pt-Co_SR}(a-c),
the reduction of the transmission due to the Pt impurity atom is relatively small compared
to the perfect ferromagnetic Co monowire. We can understand this general behavior
from the fact that the Pt $5d$-bands possess a
broader bandwidth and thereby allow transmission in the entire regime of the
spin-polarized Co $3d$ bands
(cf.~Fig.~\ref{fig:MLWFs_vs_FSWFs_BS} and Fig.~\ref{fig:Co_DOS}).

In the $s-d_{z^2}$-channel, the reduction of the transmission is similarly small for
the majority and minority spin contributions
due to the energetic alignment of the spin-split states of the Co wire with the states of the Pt impurity. 
In the majority spin channel, a significant reduction of transmission only occurs
in a region from $E_F-2.1$~eV to $E_F-0.9$~eV where the perfect conductance amounts to $G_0$. In the spin- and
orbital-decomposed density of states (DOS), Fig.~\ref{fig:Co-Pt-Co_SR}(d) and (g), we also find two resonances at
the Pt impurity located at $2.8$~eV, and $2.3$~eV below the Fermi energy in the majority
and minority spin channel, respectively. In the conductance, we observe a Fano-type line shape due to the coupling 
of the $\Delta_1$-band to these resonances.

The conductance from the $\Delta_3$-bands displays only a reduction at the bottom and top of the
band in both spin channels as the on-site energies of Co and Pt $d_{xz,yz}$-states are close in energy.
The density of states of the Pt atom, Fig.~\ref{fig:Co-Pt-Co_SR}(h), shows that the $d_{xz,yz}$
states are spin-split, carry a significant part of the Pt moment, and align well with the $\Delta_3$-bands in the Co monowire
resulting in an efficient transport channel.
The most severe change in the conductance upon introducing a Pt impurity occurs
in the $\Delta_4$-band. Here, we observe a large decrease due to scattering at the Pt impurity.
For both, the $d_{xz,yz}$ and $d_{xy, x^2-y^2}$ channels, bound states on the Pt atom can be found due to the lower
on-site potential at the Pt site. For the $\Delta_3$-symmetry there are such states at $-4.1$ eV for
the majority band and at $-3.3$ eV for both spin channels, which do not contribute to the conduction as
they are below the $\Delta_3$-band of the Co leads.
For the $\Delta_4$ symmetry there are majority states around $-2.5$ eV and a paired state at $-2$ eV with respect to the
Fermi level, not contributing to the majority channel transmission.

$\Delta_4$ electrons are only transmitted in the small overlap region around $-2.1$~eV for majority and around Fermi level
for minority states, where a very narrow band is formed in both cases. The shape of the transmission function follows
the two-peak (majority band) and three-peak (minority band) shape of the DOS of the central Pt atom.

\begin{figure}
\begin{center}
\centerline{\includegraphics[width=0.49\textwidth,angle=270]{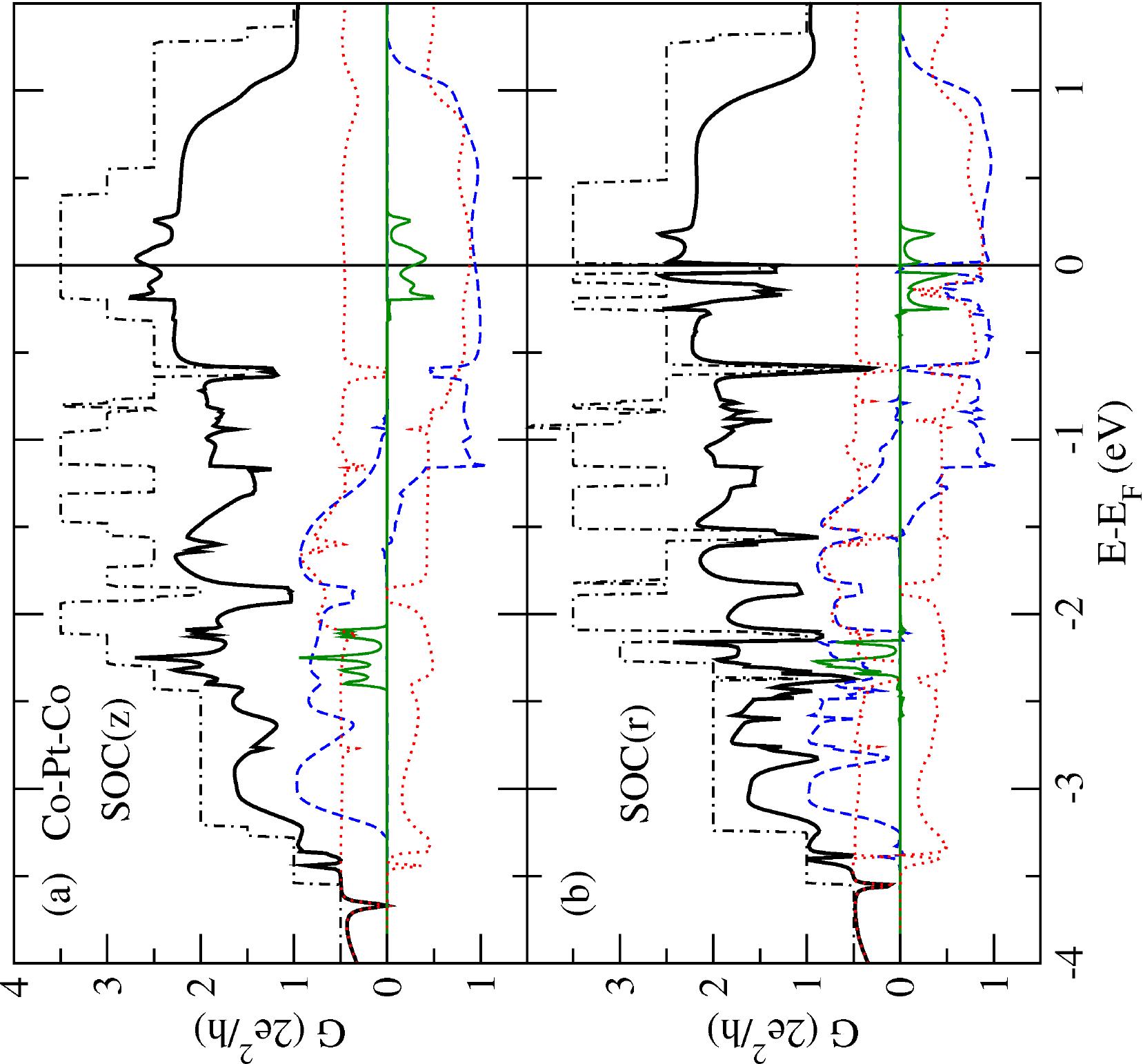}}
\caption{(color online)
Conductance including SOC for a Co monowire with a Pt impurity atom for
(a) magnetization along the wire axis and (b) perpendicular to the axis.
The decomposition of transmission into $\Delta_1$ (dotted red lines),
$\Delta_3$ (dashed blue lines) and $\Delta_4$ (solid green lines) channels
for majority spin (positive $y$-axis) and minority spins (negative $y$-axis)
shows the presence of a $\Delta_4$ minority band channel at the Fermi level.
Black dashed-dotted lines display the transmission of the perfect infinite Co
leads.
}
\label{fig:Co-Pt-Co_SOC}
\end{center}
\vspace*{-1cm}
\end{figure}

Now we turn to the effect of spin-orbit coupling on the magnetic
and transport properties of the Co-Pt-Co junction.
For the perfect Co monowire we found a magneto-crystalline anisotropy
energy (MAE), i.e.,~the difference in energy for the magnetization
in the chain axis and perpendicular to it, of $0.8$~meV per magnetic atom in favor
of an out-of-chain magnetization and orbital moments of
$0.17 \mu_B$ for out-of-chain and $0.22 \mu_B$ for in-chain
direction. Upon introducing the Pt atom, this
value is reduced to $0.5$~meV per magnetic atom, which is consistent with our
observation in the previous section for a Pt-Co-Pt junction favoring the
in-chain direction. The magnetic moment of the Pt atom is $0.36 \mu_B$
for both magnetization directions and we find similar
orbital moments of $0.09 \mu_B$ (out-of-chain) and $0.10 \mu_B$ (in-chain).
Characteristically, as in the case of the Co leads, the orbital
moments of the Co atoms adjacent to the Pt impurity are significantly
larger for the in-chain direction (reaching as much as 0.53$\mu_B$ for
the nearest Co atom), than for the out-of-chain direction (at most
0.2$\mu_B$). This means that the out-of-chain easy magnetization axis
in our scattering region is mainly due to the Co atoms.

For the transport properties including SOC, the bands with $\Delta_3$ and
$\Delta_4$ symmetry are essential. Depending on the
quantization axis defined by the magnetization direction, the degeneracy
of these bands is lifted.  In contrast to the Pt-Co-Pt system, the Co
electrodes are ferromagnetic and therefore the splitting in the
step-like conductance in the perfect Co wires changes upon
switching the quantization axis from along the chain axis to perpendicular
to it.

As can be seen from Fig.~\ref{fig:Co-Pt-Co_SOC}, changing the magnetization
direction in a perfect infinite Co chain leads to a reduction of the transmission
from 3.5 $G_0$ (along the chain) to 1.5 $G_0$ (perpendicular to the chain)
in a very small energy window around the Fermi energy, which results in a
huge value of the ballistic anisotropic magnetoresistance of 133\%.\cite{SokolovNN2007} 
In a realistic situation, however, such values of the anisotropic magnetoresistance can 
be hardly achieved, owing to the destruction of perfect conducting channels by 
imperfections, impurities and disorder.

\begin{figure}
\begin{center}
\centerline{\includegraphics[width=0.35\textwidth,angle=270]{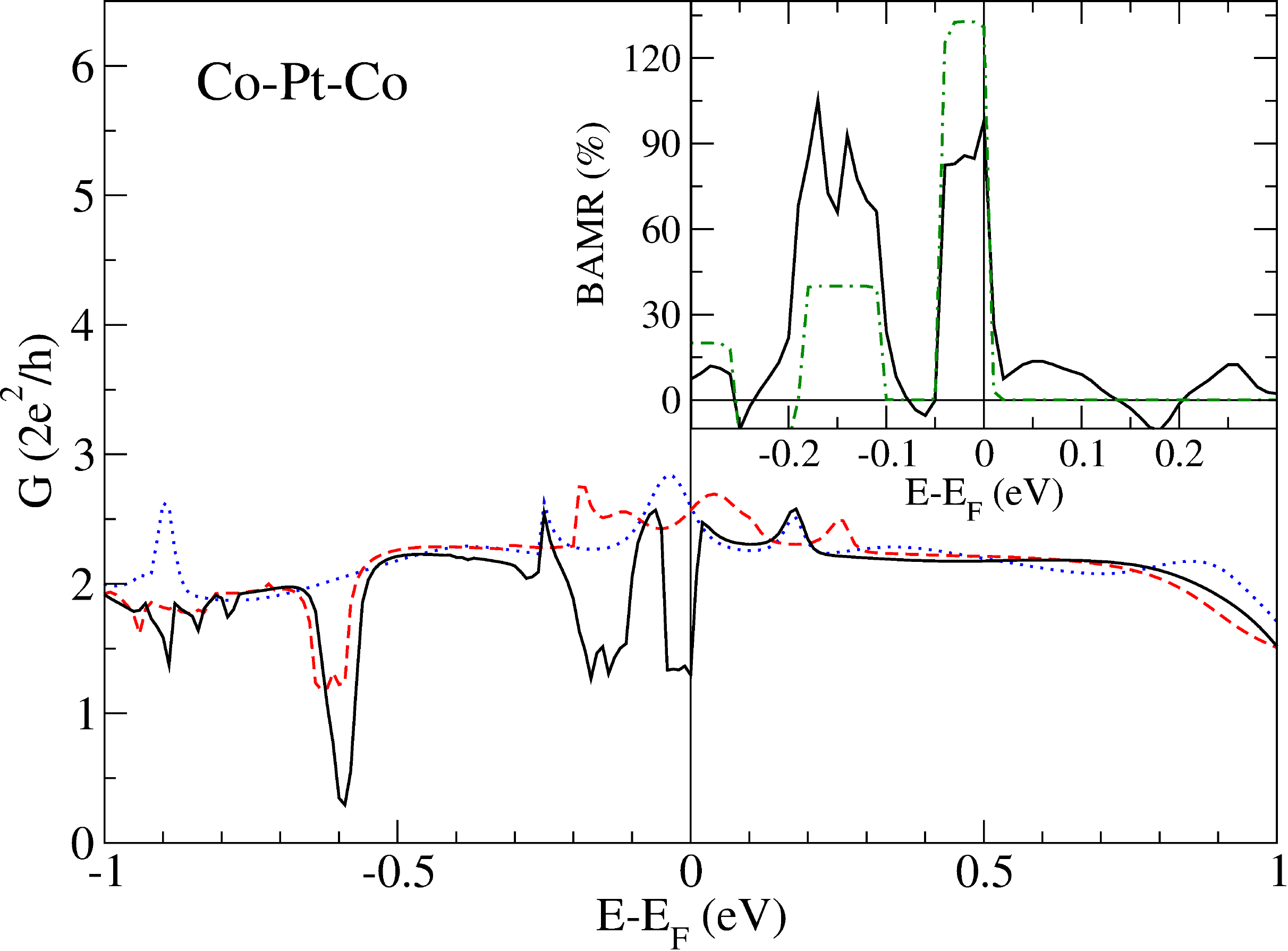}}
\caption{(color online)
Conductance around the Fermi level for a Co monowire with a Pt impurity atom in the scalar-relativistic approximation
(blue dotted line) and including SOC for a magnetization along the chain axis (dashed red line) and perpendicular to the
axis of the wire (solid black line).
The inset shows the BAMR as defined by
Eqn.~(\ref{fig:Co-Pt-Co_SR_vs_SOC})
for the Co monowire with a Pt impurity (solid black line) and for a perfect infinite Co monowire (dashed-dotted green line).
}
\label{fig:Co-Pt-Co_SR_vs_SOC}
\end{center}
\vspace*{-1cm}
\end{figure}

In the case of a Co chain with a Pt impurity, similarly to the
scalar-relativistic case, we observe a reduction by roughly a factor of two in the overall
conductance over the entire energy range, due to the less efficient coupling between the Co wire and the Pt impurity
compared to an perfect Co wire, especially for the $\Delta_3$- and $\Delta_4$-orbitals.
At the Fermi energy, we find majority and minority
spin contributions from the $\Delta_1$-band of about $0.5$ and $1.0$~$G_0$ for both magnetic directions.
Only the minority states of the other two orbital symmetries are
present due to the exchange splitting.
The minority $\Delta_3$ band contributes almost $1.0$~$G_0$
for the in-chain magnetization, while it reveals a large dip at $E_F$ for the  
out-of-chain magnetization. Accordingly, the
$\Delta_4$-band conductance also changes significantly upon switching the
magnetization direction, owing to the changes in the details of hybridization
between $\Delta_3$ and $\Delta_4$ states when the direction of the magnetization
is changed, see Fig.~\ref{fig:Co-Pt-Co_SOC}(b) and
Fig.~\ref{fig:Pt-Co-Pt_G_vs_DOS_incl_SOC} (cf.~DOS of the Co monowires for the two different magnetization directions).
These changes in the energetic
structure of $\Delta_3$ and $\Delta_4$ states lead to a large difference between
the in-chain and out-of-chain conductances, also visible for the pure Co chain
in the Fig.~\ref{fig:Co-Pt-Co_SOC}.

In Fig.~\ref{fig:Co-Pt-Co_SR_vs_SOC}, the conductance is displayed in a small energy window around the Fermi energy
for the two different magnetization directions. It is apparent that the changes arise due to the modifications of the
$\Delta_4$-band conductance between Fermi level and $-0.05$~eV, and $\Delta_3$-band conductance around $E_F$ and $-0.15$~eV
which are subject to different band mixing from spin-orbit coupling. As a result of the fine
structure of the $\Delta_4$ and $\Delta_3$ conductances (see Fig.~\ref{fig:Co-Pt-Co_SOC}(b)), the BAMR which we
obtain, shown in the inset of Fig.~\ref{fig:Co-Pt-Co_SR_vs_SOC}, displays a strong variation with energy.
Compared to the BAMR of a perfect Co MW of $133\%$ a Pt scatterer reduces this effect to $80-100\%$, which is still
considerably high. An enhanced BAMR can be found for the second peak below the Fermi Energy, where a $\Delta_4$ conduction
peak for the in-chain direction in coincidence with a $\Delta_3$ conduction depletion result in a BAMR increase from $40\%$ 
for the perfect Co MW to $60-100\%$ when a Pt scatterer is introduced.

\section{Summary}
\label{sec:Summary}
We have implemented the Landauer-B\"uttiker method to calculate the ballistic electron transport through one-dimensional
nanoscale junctions based on density functional theory calculations within the full-potential linearized augmented (FLAPW) method.
In order to apply the efficient Green's function method to calculate the conductance
we have mapped the extended Bloch states obtained from the FLAPW method to the minimal basis set of 
localized Wannier functions and constructed the Hamiltonian for the open system.
With our approach it is feasible to calculate ballistic transport through one-dimensional nanoscale systems 
including magnetism and spin-orbit coupling with the accuracy and flexibility of the FLAPW method.

We apply our method to calculate the conductance of non-magnetic Pt monowires with a single stretched
bond, including spin-orbit coupling.
Already this simple example shows the key
impact of SOC for systems containing heavy transition metals. As a second example, we considered a Co monowire and
studied the magnetoresistance upon stretching the wire at a single bond. The decomposition of the transmission into the
channels of different orbital symmetry shows the dominant contribution of $s$- and $d_{z^2}$-states as one moves from the
contact to the tunnel regime.
Finally, we studied the effect of spin-orbit scattering at an impurity atom in a monowire. We considered
two model cases: (i) a magnetic atom in a non-magnetic wire, Co in a Pt monowire, and (ii) a non-magnetic heavy element in a
ferromagnetic wire, Pt in a Co monowire. We observed for both cases a
distinct dependence of the conductance on the magnetization direction with respect to the wire axis. 

We found for a Co impurity in a Pt chain, that due to the broken cylindrical symmetry for an out-of-chain magnetization direction
the hybridization between states of different angular character and spin but with identical quantum number $j$ 
leads to scattering processes that do not conserve spin.
Those ballistic spin-scattering processes are resulting into a BAMR of 7\%.
The relatively moderate values are caused by the large background conductance from bands
originating from $s-d_{z^2}$- and $d_{xz,yz}$-states which are not modified much upon switching the magnetization. 
On the other hand, for a Pt impurity in a Co chain we find that the presence
of an impurity, although reducing somewhat the BAMR of the pure Co chain, still leads to
values of BAMR of about 100\%, which originates from hybridization between the $\Delta_3$
and $\Delta_4$ states moderated via SOC by the direction of the magnetization.


\section{Acknowledgement}

We acknowledge helpful discussions with Stefan Bl\"ugel. Funding by the DFG within the SFB677 is gratefully acknowledged.
S.H. thanks the DFG for financial support under HE3292/8-1. N.-P.W. thanks financial support from The Natural Science Foundation
of Zhejiang Province in China under Grant No. Y6100467. Y.M. and F.F. gratefully acknowledge the J\"ulich Supercomputing
Centre for computing time and funding under the HGF-YIG Programme VH-NG-513.

\appendix
\section{Computational details}

\subsection{Pt monowires}
\label{subsec:Pt_SR_CD}

Non-magnetic (NM) 6 and 12 atom supercell calculations with
an interatomic distance of $d_{\mathrm{Pt}}=4.48$ bohr and the central 
bond stretched by $\Delta= 0.0$, $0.34$, $0.72$, $1.22$, $1.82$ and $2.52$ bohr.
We applied the generalized gradient approximation (GGA) to the exchange-correlation
potential\cite{rpbe}. 
For calculations in the scalar-relativistic (SR) approximation, the
irreducible part of the 1D Brillouin zone (BZ) was sampled by 6$-$10 $k$-points depending on the
size of the supercell. 
For the 6 atom supercell, we also performed calculations including
spin-orbit coupling in second variation. For calculations with SOC the whole 1D BZ was sampled 
by 24 $k$-points. 
In all calculations, $G_{\mathrm{max}}$ was chosen to be 3.7 bohr$^{-1}$, 
which corresponds to approximately 200 basis functions per atom. 
The diameter of the cylindrical
vacuum, $D_{\rm vac}$, and the value of the in-plane auxiliary lattice constant, $\tilde{D}$,\cite{MokrousovPRB2005} 
were set to 5.0 and 7.3~bohr, respectively.

For the conductance calculations we applied the locking technique to a
perfect monowire to describe the semi-infinite leads (see Sec.~\ref{subsection:locking}).
In the SR approximation FSWFs and MLWFs were generated on a mesh of 16 $k$-points in the
whole BZ starting from one $4s$- and 5 $3d$-orbitals per atom in the
supercell, 
based on solutions of the radial equation of the
first-principles potential as trial functions.
In the calculations including SOC, MLWFs were generated
on a 24 $k$-point mesh in the whole BZ based on 2 radial $4s$- and 10 radial
$3d$-orbitals per atom,
based on solutions of the radial equation of the
first-principles potential as trial functions, due to the coupled spin channels. The energy bands were disentangled
using the procedure described in Ref.~[\onlinecite{disentanglement}]. For the SR
calculations, the lowest 80 eigenstates are needed
for 72 WFs for the 12 atom supercell and the lowest 44 eigenvalues per $k$-point for
36 WFs for the 6 atom supercell calculations. With SOC the lowest $80$ eigenstates 
per $k$-point for 72 WFs were used.

\subsection{Co monowires}
\label{Co-compdet}

Calculations with a lattice constant of $d_{\mathrm{Co}}=4.15$ bohr and a central
stretched bond with stretching $\Delta= 0.0$, $0.45$, $1.05$, $1.85$ and $2.85$ bohr.
Two collinear magnetic configurations of the Co monowire are considered, 
parallel or antiparallel alignment of the Co spins on the left and on the 
right sides of the gap, described by performing two calculations: 
A 8 atom supercell constructed from two 4-spin blocks separated
by a gap and aligned in parallel (up), while in order to mimic the 
antiparallel alignment, we considered 16 atoms in the supercell with 4-spin (up),
8-spin (down) and 4-spin (up) blocks, separated by two gaps with the spins
antiparallel to each other at each side of the gap. 

The perfect lead ferromagnetic Co monowire was calculated with 24 $k$-points
in the whole BZ, using the $G_{\mathrm{max}}$ of 4.1~bohr$^{-1}$ ($\approx$
220 basis functions per atom).
For both 8 and 16 atom supercell calculations the irreducible part of the 1D
Brillouin zone (BZ) was sampled by 8 $k$-points and $G_{\mathrm{max}}$ was
chosen to be 3.7 bohr$^{-1}$ resulting in approximately 210 basis functions 
per atom. The vacuum parameters $D_{\rm vac}$ and $\tilde{D}$ constituted 
4.3 and 6.6~bohr, respectively, in all cases. The exchange-correlation
potential was treated within the GGA\cite{rpbe}.  For all quantum conductance
calculations the locking technique (see Sec. \ref{subsection:locking}) to a
perfect FM Co monowire was used.  As trial orbitals for the FSWFs 6
$s$- and $d$-orbitals per atom and spin in the supercell were used, 
based on solutions of the radial equation of the
first-principles potential.
For the disentanglement procedure\cite{disentanglement}
the lowest 58 (110) eigenstates per $k$-point were used to obtain the 48 (96) WFs
in the 8 (16) atom supercell calculation.

\subsection{Scattering on impurities}
\label{subsec:Co-Pt-Co_CD}

A 9 atom supercell were used for the scattering region consisting of one impurity atom (Pt or Co) 
and 4 monowire atoms (Co or Pt) on both sides. 
The interatomic distance was chosen as $d_{\mathrm{Co}}=4.15$ bohr for
the Co monowire with a Pt impurity and as $d_{\mathrm{Pt}}=4.48$ bohr for the Pt monowire
with a Co impurity. 
The exchange-correlation potential was treated within the GGA\cite{rpbe}
and SOC was included in second variation. All calculations 
were performed in the scalar-relativistic (SR) approximation and for two different
directions of the magnetization with SOC, along the chain axis and perpendicular to it.
The 1D Brillouin zone (BZ) was sampled by 16 $k$-points and $G_{\mathrm{max}}$ was set to
3.9~$\mathrm{bohr}^{-1}$ resulting in approximately 175 (190) basis functions per atom
for the Co (Pt) monowire with a Pt (Co) impurity. 
For the case of an isolated Pt impurity, 
the leads were described by a Co monowire in a 3 atom unit cell in either
the scalar-relativistic (SR) approximation or including SOC for the magnetization direction
along the wire axis or perpendicular to it. The BZ was sampled by 24 $k$-points and
$G_{\mathrm{max}}$ was set to 4.1~$\mathrm{bohr}^{-1}$, resulting in approximately 210 basis
functions per atom.  For Pt monowire with a Co impurity, the lead's electronic
structure was obtained from calculations of perfect Pt monowires.
The vacuum parameters for all cases constituted 4.3 and 6.6~bohr
for $D_{\rm vac}$ and $\tilde{D}$, respectively.

For all quantum conductance calculations the locking technique 
wa used and the third nearest-neighbor approximation were employed.
In  the SR case FSWFs were generated on a 16 $k$-point mesh in 
the whole 1D-BZ with 1 $s$- and 5
$d$-orbitals per atom and spin,
based on solutions of the radial equation of the
first-principles potential.
For the disentanglement procedure\cite{disentanglement}
the lowest 64 (62) eigenvalues per $k$-point for 54 (54) WFs for Pt (Co) impurities in Co (Pt) monowires were considered.
The Pt and Co lead WFs were constructed as described in Sec.~\ref{subsec:Pt_SR_CD}
and Sec.~\ref{Co-compdet} in this case.
With SOC the FSWFs were generated on a 16 $k$-point mesh in the whole 
1D-BZ with 2 $s$- and 10 $d$-orbitals per atom,
based on solutions of the radial equation of the
first-principles potential.
For disentanglement\cite{disentanglement}
the lowest $116$ eigenstates per $k$-point for $108$ WFs were used.
The WFs for the semi-infinite Co leads were generated on a 24 $k$-point mesh with the 
same trial functions as those used for the atoms inside the scattering region, while 
for disentanglement the lowest $26$ eigenvalues per $k$-point for $18$ WFs per spin 
(SR) and the lowest $44$ eigenvalues per $k$-point for $36$ WFs (SOC) were used. The Pt lead WFs were
constructed as described in Sec.~\ref{subsec:Pt_SR_CD}.


\end{document}